\documentclass[journal]{IEEEtran}
\usepackage[T1]{fontenc}
\usepackage{amsmath,amssymb,mathtools,amsthm}
\usepackage{graphicx}
\usepackage{bm}
\usepackage{booktabs}
\usepackage{url}
\usepackage{xcolor}
\definecolor{linkgreen}{RGB}{0,110,60}
\usepackage[colorlinks=true,linkcolor=red,citecolor=blue,urlcolor=blue]{hyperref}
\usepackage{microtype}

\newcommand{\maybeincludegraphics}[2][]{%
  \IfFileExists{#2}{\includegraphics[#1]{#2}}{%
    \fbox{\parbox[c][1.0in][c]{0.88\linewidth}{\centering Figure file not supplied:\\ \texttt{\detokenize{#2}}}}}}
\newtheoremstyle{clean}{7pt}{7pt}{\itshape}{}{\normalfont}{.}{.5em}{}
\theoremstyle{clean}
\newtheorem{theorem}{Theorem}
\newtheorem{lemma}{Lemma}
\newtheorem{proposition}{Proposition}
\newtheorem{corollary}{Corollary}
\theoremstyle{definition}

\newcommand{\R}{\mathbf R}
\newcommand{\B}{\mathbf B}
\newcommand{\I}{\mathbf I}
\newcommand{\CN}{\mathcal{CN}}
\newcommand{\E}{\mathbb E}
\newcommand{\Prb}{\mathbb P}
\newcommand{\Nzero}{\mathbb N_0}

\newcommand{\diag}{\operatorname{diag}}

\title{Fast Fluid Antenna Multiple Access: Exact Analysis and Fixed-Aperture Asymptotics}

\author{Gustavo Fraidenraich,
Thales Wulfert Cabral,
Farshad Rostamighadi, \IEEEmembership{Senior Member,~IEEE},\\ 
Kai-Kit Wong, \IEEEmembership{Fellow,~IEEE}, and 
Kin-Fai Tong, \IEEEmembership{Fellow,~IEEE}
\vspace{-5mm}

\thanks{G. Fraidenraich and T. W. Cabral are with the School of Electrical and Computer Engineering, University of Campinas, Campinas, Brazil.}\thanks{F. Rostamighadi and K. K. Wong are with the Department of Electronic and Electrical Engineering, University College London, London, U.K. and K. K. Wong is also affiliated with the Department of Electronic Engineering, Kyung Hee University, Yongin-si, Gyeonggi-do 17104, Korea
(e-mail: \{f.rostamighadi, kai-kit.wong\}@ucl.ac.uk).}
\thanks{K. F. Tong is affiliated with the School of Science and Technology, Hong Kong Metropolitan University, Hong Kong SAR, China (e-mail: ktong@hkmu.edu.hk).}
}
\begin{document}
\maketitle
\begin{abstract}
In this paper, we investigate the interference-limited outage performance of fast fluid antenna multiple access (FAMA), in which the receive port is selected according to the instantaneous symbol-based signal-to-interference energy ratio. For any positive-definite spatial covariance matrix and an arbitrary finite number of candidate ports, we show that constant-modulus multiuser interference admits an exact reduction to the comparison of two independent correlated exponential random vectors. This representation leads to a nonnegative double multi-index series for the outage probability, a recursive coefficient evaluation, and a certified truncation-error bound. We then study the distinct asymptotic regime where the number of candidate ports goes to infinity while the physical aperture remains unchanged, assuming exponential spatial correlation. For this model, we derive an explicit finite-port series requiring no numerical integration, and characterize continuous port selection through a convergent one-dimensional series for arbitrary thresholds. The analysis further yields an exact expression relating the maximum supportable number of users to the antenna aperture size and target outage probability. Numerical results validate the finite-port and asymptotic analyses, and illustrate that given sufficient aperture size, a user utilizing fast FAMA can overcome a massive number of interferers for decent communication performance.
\end{abstract}


\vspace{-2mm}
\section{Introduction}\label{sec:intro}
\IEEEPARstart{M}{ultiple access} is arguably one of the most fundamental problems in wireless communications, since it determines how limited spectral resources are shared among multiple simultaneous users. Classical multiple-access schemes such as time-division multiple access (TDMA) and frequency-division multiple access (FDMA) provide simple and robust mechanisms for orthogonal resource sharing and remain widely used today \cite{Sekimoto-1968,Puente-1971}. However, their spectral efficiency is inherently limited because users are separated in time or frequency. Over the past two decades, the focus has increasingly shifted towards spatial-division multiple access (SDMA) \cite{Ko-1983}, enabled by multiple-input multiple-output (MIMO) technology, where users are separated in the spatial domain and can therefore occupy the same time-frequency resources. Multiuser MIMO has consequently become one of the defining technologies of modern wireless communications and remains central to both current and future cellular systems \cite{Wong-2000,Wong-2002,Vishwanath-2003,Spencer-2004}.

Increasingly, however, the scalability of multiuser MIMO raises serious concerns. As the number of simultaneously served users and antenna elements grows, the complexity of channel acquisition and precoder optimization can become substantial. Although massive MIMO was originally motivated in part by the prospect of using very large antenna arrays to simplify multiuser processing, practical systems cannot increase the number of active antennas without limit \cite{Marzetta-2010}. Beyond the associated hardware, RF-chain and power-consumption costs, aggressive multiuser precoding can also create demanding power-amplifier requirements, including increased signal-envelope fluctuations and peak-to-average power ratio \cite{Sohrabi-2016}. Consequently, the regime in which very simple precoding becomes near-optimal is difficult to realize in practice at truly large scale. Multiuser MIMO therefore remains our most dependable framework for spatial multiple access, but its future scaling path is far from obvious. This motivates the search for fundamentally new ways of creating and exploiting spatial degrees of freedom without relying solely on ever-larger conventional antenna arrays.

It is against this backdrop that the fluid antenna system (FAS) concept has emerged in recent years as a new approach to exploiting spatial degrees of freedom \cite{Wong-fas2020cl,wong2020fluid}. FAS is a hardware-agnostic system paradigm in which the antenna is treated as a reconfigurable physical-layer resource, allowing the effective antenna configuration to be adapted within a prescribed spatial region through position, shape, radiation-pattern, or other forms of reconfigurability \cite{new2025tutorial,Lu-2025,hong2025contemporary,New-2026jsac,Wu-2026}. One particularly distinctive application is fluid antenna multiple access (FAMA), where this reconfigurability is exploited directly for multiuser interference management \cite{Hong-fama2026}. By selecting favourable antenna states at the receiver, FAMA can suppress aggregate interference in the spatial domain and therefore provide a fundamentally different route to multiple access that complements, and even reduces the reliance on conventional transmitter-side multiuser precoding.

Existing FAMA schemes can be broadly grouped into: 
\begin{itemize}
\item Fast FAMA \cite{wong2022fama,wong2023fast}, 
\item Slow FAMA utilizing single-port selection \cite{wong2023slow,xu2024revisiting,Espinosa-2024} and multi-port selection and combining \cite{wong2024cuma}, and 
\item Coded FAMA \cite{hong2025coded,hong20255gcoded,waqar2025turbocharging}, 
\end{itemize}
according to how antenna reconfigurability is exploited and over what timescale the receiver adapts. Fast FAMA performs rapid antenna-state or port selection based on instantaneous interference conditions, with the objective of identifying spatial states in which the per-symbol aggregate multiuser interference is naturally suppressed. Slow FAMA operates over a longer timescale, attempting to maximize the signal-to-interference plus noise ratio (SINR) averaged over the information symbols when selecting favourable antenna configurations with lower implementation and estimation overhead. Coded FAMA, on the other hand, combines antenna reconfigurability with channel coding or coded transmission principles. 

Slow FAMA is generally regarded as the more practical operating mode, whereas fast FAMA has often been viewed as difficult to realize despite its remarkable multiple-access capability, primarily because practical mechanisms for identifying the optimal antenna port on a symbol-by-symbol basis remain unclear. Consequently, most existing performance analyses of FAMA have focused on slow FAMA \cite{wong2023slow,xu2024revisiting,Espinosa-2024}. More recently, nonetheless, \cite{waqar2026fast,waqar2026turbo} showed that a learning-based approach can approximate fast-FAMA port selection, renewing interest in understanding its fundamental performance limits. To date, the only analytical performance studies of fast FAMA appear in \cite{wong2022fama,Moura-2026}. That analysis, however, relies on a highly simplified spatial-correlation model adopted for mathematical tractability, leaving the performance of fast FAMA under more realistic spatial dependence largely unresolved.

The main contributions are as follows:
\begin{enumerate}
\item For any arbitrary finite number of ports, $N$, and positive-definite spatial correlation matrix, $\R$, we first derive a nonnegative double multi-index outage series in scalar incomplete beta functions, with a coefficient recursion and certified truncation error, applicable for characterizing fast FAMA systems. The constant-modulus interference reduction is exact; small-threshold and two- and three-port expressions allow analytical checks.
\item Under exponential spatial correlation, we derive explicit finite-grid mixture weights and the exact limit as $N\to\infty$ with a given aperture, including a convergent spectral series for general signal-to-interference ratio (SIR) thresholds, and a closed form at unit normalized threshold. 
\item Moreover, we link the maximum supportable user count to aperture, SIR target, and outage requirement through exact numerical inversion, an explicit large-load approximation, and a conservative explicit guarantee.
\end{enumerate}
%The results retain spatial correlation and concern ideal port selection and per-user outage.

\vspace{-2mm}
\section{Fast FAMA System Model}\label{sec:model}
Consider a fast-FAMA system where each antenna of a central transmitter (e.g., base station (BS)) is responsible for sending information symbol $\tilde{s}_\iota$ to a user, say $\iota$, on the same channel use, for $\iota=0,1,\dots,U-1$. For simplicity, we assume $|\tilde{s}_\iota|=1, \forall \iota$. Each user is equipped with a FAS of $N$ candidate ports and a linear aperture of $W\lambda_{\rm c}$ where $\lambda_{\rm c}$ is the carrier wavelength and $W$ is the normalized length. Considering user $0$ as the desired user, the received signal at port $i$ is%\footnote{We assume that interference dominates and noise is thus ignored.}
\begin{equation}\label{eq:basic_model}
\tilde r_i=h_i\tilde{s}_0+\sum_{\iota=1}^{L}g_{\iota,i}\tilde{s}_\iota\equiv h_i\tilde{s}_0+z_i,
\end{equation}
where $L\triangleq U-1$, $ z_i\triangleq\sum_{\iota=1}^{L}g_{\iota,i}\tilde{s}_\iota$, $h_i$ denotes the desired user's channel at port $i$ and $g_{\iota,i}$ represents the interfering channel from user $\iota\ne 0$ seen at port $i$. The channels at different ports are correlated so that $\mathbf h=(h_1,\dots,h_N)^T\sim\CN(\mathbf0,\R)$, while the interfering channel vectors are mutually independent and independent of the desired channel, with $\mathbf g_\iota=(g_{\iota,1},\dots,g_{\iota,N})^T\sim\CN(\mathbf0,\R)$. Here $\CN(\mathbf0,\R)$ is a zero-mean proper complex Gaussian distribution with covariance $\R(\succ0)$, meaning that $\R$ is Hermitian positive definite, and $[\R]_{ii}=1$ normalizes the channel power at every port. Note that in (\ref{eq:basic_model}), noise is neglected as $L$ is considered very large.

%\begin{figure*}[!t]
%\centering
%\maybeincludegraphics[width=\textwidth]{figures/fast_fama_system_model.pdf}
%\caption{System model of fast FAMA with one desired user, $L=U-1$ interferers, and $N$ spatially correlated receive ports. The receiver selects the port with the highest instantaneous SIR, accounting for the complex aggregate interference.}\label{fig:system_model}
%\end{figure*}

In fast FAMA, user $0$ estimates the symbol-based SIRs at all ports \cite{waqar2026fast,waqar2026turbo} and selects the optimal port maximizing the symbol-based SIR \cite{wong2022fama,wong2023fast}. That is,
\begin{equation}\label{eq:system}
\gamma_{\max}=\max_{1\le i\le N}\frac{|h_i|^2}{|z_i|^2}.
\end{equation}
A key performance metric is outage probability which measures the probability that a given performance threshold $t>0$ is violated. Mathematically, that is $P_{\rm out}(t)=\Prb(\gamma_{\max}<t)$. %This is an ensemble probability at a symbol instance, not a statement that consecutive symbols are independent in a fading block.

\begin{proposition}[Exact normalized model]\label{prop:reduction}
For $U\ge2$, let $q=Lt$. The outage probability can be characterized by
\begin{equation}\label{eq:normalized}
\left\{\begin{aligned}
P_{\rm out}(t)&=F_{N,\R}(q),\\
F_{N,\R}(q)&=\Prb(X_i<qY_i,\ i=1,\ldots,N),
\end{aligned}\right.
\end{equation}
where $X_i\triangleq |h_i|^2$, $Y_i\triangleq |g_i|^2$, and $\mathbf h=(h_1,\dots, h_N), \mathbf g=(g_1,\dots, g_N)$ are independent $\CN(\mathbf0,\R)$ vectors.
\end{proposition}

\begin{proof}
Conditioned on the symbols, $\mathbf z=(z_1,\dots,z_N)^T$ is proper complex Gaussian with covariance $\sum_{\iota=1}^{L}|\tilde{s}_\iota|^2\R=L\R$. Consequently, for $U\ge2$, $\mathbf z/\sqrt L\sim\CN(\mathbf0,\R)$, independent of $\mathbf h$, and substitution into \eqref{eq:system} proves the result.
\end{proof}

Under rich scattering, we have the Jakes' model so that
\begin{equation}\label{eq:jakes}
[\R]_{ij}=J_0(2\pi|i-j|\Delta),
\end{equation}
in which $\Delta\triangleq\frac{W}{N-1}$ is the port spacing for $N\ge2$, and $J_0(\cdot)$ denotes the zeroth-order Bessel function of the first kind.
%The general results also admit nonuniform positions and complex Hermitian covariance matrices.

For non-constant-modulus interfering symbols, the same argument gives the conditional extension
\begin{equation}\label{eq:randomenergy}
P_{\rm out}(t)=\E_{\mathcal E_{\rm sym}}[F_{N,\R}(t\mathcal E_{\rm sym})],
\end{equation}
where $\mathcal E_{\rm sym}=\sum_{\iota=1}^{L}|\tilde{s}_\iota|^2$ is the aggregate symbol energy of the interfering users, provided the numerator remains $|h_i|^2$. %Replacing $S$ by its mean is not generally exact. Unequal user covariance matrices and additive noise are outside the identical-covariance theorem below.

\vspace{-2mm}
\section{Exact Outage Probability for Finite $N$}\label{sec:mixture}\label{sec:outage}
\subsection{The General Expression}
We evaluate the coupled ratio event in \eqref{eq:normalized} using a Gamma-mixture representation of correlated Gaussian powers \cite{royen1991,morales2011}. Unlike conventional antenna selection combining, this event compares two independent correlated vectors. Let $\zeta_i\ge0$, $i=1,\ldots,N$, be the real Laplace-transform arguments, and set $\mathbf S_{\rm L}\triangleq\diag(\zeta_1,\ldots,\zeta_N)$. Unlike the transmitted symbols $\tilde{s}_\iota$ in Section~\ref{sec:model}, the variables $\zeta_i$ are deterministic transform parameters. In the Laplace domain, the Gaussian quadratic-form identity of ${\bf X}=(X_1,\dots,X_N)$ gives \cite{turin1960}
\begin{equation}
 \mathcal L_{\mathbf X}(\mathbf S_{\rm L})
 =\E_{\bf h}\left[e^{-\mathbf h^H\mathbf S_{\rm L}\mathbf h}\right]
 =\frac{1}{\det(\I+\R\mathbf S_{\rm L})}.
 \label{eq:joint-laplace}
\end{equation}

Choose $0<a\le\lambda_{\min}(\R)$ and define
\begin{equation}\label{eq:B}
\B\triangleq\I-a\R^{-1},~\mbox{and }C\triangleq\frac{a^N}{\det\R}=\det({\bf I}-{\bf B}).
\end{equation}
Note that the chosen range of $a$ ensures that $\B$ is positive semidefinite, with all the eigenvalues smaller than one. This property allows the determinant below to be expanded into a convergent series with nonnegative coefficients.

Introducing $\upsilon_i=(1+a \zeta_i)^{-1}$, $\boldsymbol\upsilon=(\upsilon_1,\ldots,\upsilon_N)$ and $\mathbf V=\diag(\upsilon_1,\ldots,\upsilon_N)$, we obtain from \eqref{eq:joint-laplace} that 
\begin{equation}\label{eq:laplace-transformed}
\mathcal L_{\mathbf X}(\mathbf S_{\rm L})=\frac{C}{\det(\I-\B\mathbf V)}\prod_{i=1}^N\upsilon_i.
\end{equation}
Now, expanding the remaining determinant into a multivariate power series accordingly gives
\begin{equation}\label{eq:pgf}
G(\upsilon_1,\ldots,\upsilon_N)\triangleq\frac{C}{\det(\I-\B\mathbf V)}=\sum_{\mathbf m\ge\mathbf0}w_{\mathbf m}\boldsymbol\upsilon^{\mathbf m},
\end{equation}
where $\mathbf m=(m_1,\ldots,m_N)$, $\boldsymbol\upsilon^{\mathbf m}=\prod_i \upsilon_i^{m_i}$, and $\mathbf m\ge\mathbf0$ means $m_i\ge0$ for every $i$. The coefficients $w_{\mathbf m}$ are introduced in (\ref{eq:wm}) of Appendix \ref{appA-1} and their expressions will be presented later in (\ref{eq:recursion}). Substituting (\ref{eq:pgf}) into \eqref{eq:laplace-transformed} yields
\begin{equation}\label{eq:laplace-gamma}
\mathcal L_{\mathbf X}(\mathbf S_{\rm L})=\sum_{\mathbf m\ge\mathbf0}w_{\mathbf m}\prod_{i=1}^N(1+a \zeta_i)^{-(m_i+1)}.
\end{equation}

Each product in \eqref{eq:laplace-gamma} is the Laplace transform of independent Gamma variables with shape parameters $m_i+1$ and common scale $a$. By uniqueness of the Laplace transform, the joint density is therefore given by
\begin{equation}\label{eq:gamma}
f_{\mathbf X}(\mathbf x)=\sum_{\mathbf m\ge\mathbf0} w_{\mathbf m}\prod_{i=1}^N\frac{x_i^{m_i}e^{-x_i/a}}{a^{m_i+1}m_i!},~\mathbf x\ge\mathbf0.
\end{equation}
The coefficients $w_{\mathbf m}$ are nonnegative and satisfy $\sum_{\mathbf m\ge\mathbf0}w_{\mathbf m}=1$, as demonstrated in Appendix \ref{appA-1}. Thus, spatial correlation is fully retained through the mixture coefficients, while the Gamma variables are independent within each term.

\begin{theorem}[Exact outage probability]\label{thm:main}
For arbitrary $N$, any positive-definite spatial covariance matrix $\R$, and $q>0$,
\begin{equation}\label{eq:main}
F_{N,\R}(q)=\sum_{\mathbf m,\mathbf n\ge\mathbf0} w_{\mathbf m}w_{\mathbf n}\prod_{i=1}^N I_p(m_i+1,n_i+1).
\end{equation}
All terms are nonnegative and the series converges absolutely.
\end{theorem}

\begin{proof}
Using the probability density function (PDF) in \eqref{eq:gamma} for the two independent vectors, ${\bf X}$ and ${\bf Y}$, and substituting them into \eqref{eq:normalized} gives
\begin{equation}\label{eq:outage-gamma}
F_{N,\R}(q)=\sum_{\mathbf m,\mathbf n\ge\mathbf0} w_{\mathbf m}w_{\mathbf n}\prod_{i=1}^N\Prb(G_{m_i+1}<qG'_{n_i+1}),
\end{equation}
where $G_k$ and $G'_\ell$ are independent Gamma variables with shape parameters $k$ and $\ell$ and common scale $a$. Using the standard relationship between independent Gamma variables and the Beta distribution, the ratio $G_k/(G_k+G'_\ell)$ follows a Beta$(k,\ell)$ distribution. Consequently, we have
\begin{equation}\label{eq:beta-step}
\Prb(G_k<qG'_\ell)=I_p(k,\ell),~\mbox{with }p=\frac{q}{1+q},
\end{equation}
where $I_p(k,\ell)$ denotes the regularized incomplete beta function. Applying (\ref{eq:beta-step}) in (\ref{eq:outage-gamma}) gives the desired result (\ref{eq:main}).
\end{proof}

%For integer shape parameters, the regularized incomplete beta function can be expressed as the finite sum
%\begin{multline}\label{eq:finitebeta}
%I_p(m+1,n+1)=\sum_{j=m+1}^{m+n+1}\binom{m+n+1}{j}\\
%\times p^j(1-p)^{m+n+1-j}.
%\end{multline}
%Therefore, \eqref{eq:main} provides an exact analytical representation of the outage probability in fast FAMA systems for arbitrary finite $N$ and general positive-definite spatial correlation, without requiring multidimensional numerical integration.

\vspace{-2mm}
\subsection{Evaluation and Truncation}
For practical evaluation, let $\B_{{\cal S},{\cal S}}$ be a principal submatrix of $\B$ indexed by ${\cal S}\subseteq\{1,\ldots,N\}$, and let $\mathbf1_{\cal S}$ be its indicator vector. Then the principal-minor expansion gives
\begin{equation}
\det(\I-\B\mathbf V)=\sum_{{\cal S}\subseteq\{1,\ldots,N\}}(-1)^{|{\cal S}|}\det(\B_{{\cal S},{\cal S}})\prod_{i\in {\cal S}}\upsilon_i.
\end{equation}
Multiplying the right-hand side of the above onto both sides of \eqref{eq:pgf} then gives rise to
\begin{align}
C&\stackrel{(a)}{=}\left(\sum_{\mathbf m\ge\mathbf0}w_{\mathbf m}\boldsymbol\upsilon^{\mathbf m}\right)\left(\sum_{{\cal S}}(-1)^{|{\cal S}|}\det(\B_{{\cal S},{\cal S}})\prod_{i\in {\cal S}}\upsilon_i\right)\notag\\
&\stackrel{(b)}{=}\left(\sum_{\mathbf m\ge\mathbf0}w_{\mathbf m}\boldsymbol\upsilon^{\mathbf m}\right)\left(1+\sum_{{\cal S}\ne\varnothing}(-1)^{|{\cal S}|}\det(\B_{{\cal S},{\cal S}})\prod_{i\in {\cal S}}\upsilon_i\right)\notag\\
&\stackrel{(c)}{=}\sum_{\mathbf m\ge\mathbf0}w_{\mathbf m}\boldsymbol\upsilon^{\mathbf m}+\sum_{\mathbf m\ge\mathbf0}w_{\mathbf m}\boldsymbol\upsilon^{\mathbf m}\sum_{{\cal S}\ne\varnothing}(-1)^{|{\cal S}|}\det(\B_{{\cal S},{\cal S}})\prod_{i\in {\cal S}}\upsilon_i.\label{eq:compcoeff}
\end{align}
Comparing coefficients associated with $\boldsymbol\upsilon^{\mathbf m}$ in \eqref{eq:compcoeff} yields
\begin{subequations}\label{eq:recursion}
\begin{align}
w_{\mathbf0}&=C,\\
w_{\mathbf m}&=\sum_{\substack{\varnothing\ne {\cal S}\subseteq\{1,\ldots,N\}\\
\mathbf m\ge\mathbf1_{\cal S}}}(-1)^{|{\cal S}|+1}\det(\B_{{\cal S},{\cal S}})w_{\mathbf m-\mathbf1_{\cal S}}.
\end{align}
\end{subequations}
%The determinant expansion underlying this recursion is given in Appendix \ref{appA-1}. Note that although the weights are nonnegative, the alternating recursion may require increased numerical precision for strongly correlated ports.

Let $F_{N,\R}^{[K]}(q)$ be \eqref{eq:main} restricted to $|\mathbf m|,|\mathbf n|\le K$ and $\mathcal A_K\triangleq\sum_{|\mathbf m|\le K}w_{\mathbf m}$. As every factor $I_p$ lies between zero and one, the omitted probability is bounded by the total omitted weight:
\begin{equation}\label{eq:certificate}
F_{N,\R}^{[K]}(q)\le F_{N,\R}(q)\le\min\{1,F_{N,\R}^{[K]}(q)+1-\mathcal A_K^2\}.
\end{equation}

The following proposition addresses the particular problem of how to compute the retained weight, $\mathcal A_K$.

\begin{proposition}
The retained weight, $\mathcal A_K$, can be computed via the eigenvalues $\beta_j$ of $\B$ with the generating function
\begin{equation}\label{eq:spectral}
\sum_{k\ge0}a_k\upsilon^k=\prod_{j=1}^N \frac{1-\beta_j}{1-\beta_j\upsilon},~\mbox{with }\mathcal A_K=\sum_{k=0}^K a_k.
\end{equation}
\end{proposition}

\begin{proof}
Setting $\upsilon_1=\cdots=\upsilon_N=\upsilon$ in (\ref{eq:pgf}), we have
\begin{equation}
G(\upsilon,\ldots,\upsilon)=\sum_{\mathbf m\geq\mathbf0}w_{\mathbf m}\upsilon^{|\mathbf m|}=\sum_{k=0}^{\infty}a_k\upsilon^k,
\end{equation}
where $a_k=\sum_{|\mathbf m|=k}w_{\mathbf m}$ is the sum of all the multivariate mixture weights whose indices have total degree \(k\).

Alternatively, using the eigenvalues \(\beta_j\) of \(\mathbf B\), we have
\begin{align}\label{eq:gpoly-v}
G(\upsilon,\ldots,\upsilon)&=\frac{C}{\det(\mathbf I-\upsilon\mathbf B)}\notag\\
&=\frac{\det(\mathbf I-\mathbf B)}{\det(\mathbf I-\upsilon\mathbf B)}=\prod_{j=1}^N\frac{1-\beta_j}{1-\beta_j\upsilon},
\end{align}
which gives the generating function for the coefficients $a_k$.
\end{proof}

Each factor in (\ref{eq:gpoly-v}) is the probability generating function of a geometric random variable with parameter \(1-\beta_j\). Thus, \(a_k\) can also be interpreted as the probability that the sum of these independent geometric random variables equals $k$. However, note that the geometric random variables are only an auxiliary representation of the total-degree distribution, and we do not assume that the individual components of the original mixture index are independent. Finally, we can obtain
\begin{equation}
{\cal A}_K=\sum_{|\mathbf m|\leq K}w_{\mathbf m}=\sum_{k=0}^K a_k=\Pr\left(\sum_{\ell=1}^N J_\ell\leq K\right),
\end{equation}
where \(J_\ell\) are independent geometric random variables.

\vspace{-2mm}
\section{Asymptotic Case: $N\to\infty$ at Fixed Aperture and Exponential Spatial Correlation}\label{sec:exp-dense-fast}
We now consider the limiting regime where the number of candidate ports increases without bound while the physical aperture $W$ remains fixed. The port spacing is $\Delta=W/(N-1)$, so adjacent-port correlation approaches one as $N\to\infty$. %This limit cannot be obtained by assuming independent ports.

%The main motivation is mathematical tractability, but there is also a practical justification. The exponential correlation model is widely used in spatially correlated MIMO channels and provides a simple characterization of spatial decorrelation through a single parameter.
%
%Unlike Jakes? model, which assumes isotropic scattering, the exponential model serves as an alternative statistical description of spatial dependence. More importantly, it leads to a Gaussian Markov process, which allows us to obtain the analytical results in Section V.
%
%We should make it clear that we are not claiming that exponential correlation is a universally accurate approximation of Jakes? model. Rather, we adopt it as a practically relevant alternative that enables an exact continuous-aperture analysis. I think we should explicitly explain this distinction in Section V.

Instead of Jakes model, we adopt the exponential spatial-correlation model not only for analytical tractability but also because it provides a widely used representation of finite spatial correlation, accounting for the correlation decay across the aperture \cite{Lim-2017,Rodrigues-2019}. Unlike the Jakes model, whose oscillatory correlation structure follows from the specific assumption of isotropic rich scattering, the exponential model enables the effect of correlation strength to be isolated from the detailed angular characteristics of the propagation environment. It thus provides a useful benchmark for establishing fundamental system behaviour. Moreover, exponential correlation has recently been shown to provide an effective analytical approximation to Jakes-type spatial covariance in FAS settings \cite{Lee-2025}.

Specifically, we use the unit-decay exponential covariance
\begin{equation}\label{eq:exp-covariance}
{\rm COV}(\varsigma,\varsigma')=e^{-|\varsigma-\varsigma'|},
\end{equation}
for two independent, stationary, unit-variance complex Gaussian channel fields $h(\varsigma)$ and $g(\varsigma)$. As in Section~\ref{sec:model}, we set $X(\varsigma)=|h(\varsigma)|^2$, $Y(\varsigma)=|g(\varsigma)|^2$, and $q=(U-1)t$. On the uniform spatial sampling grid $\varsigma_i=i\Delta$, the outage event and its limit as $N\to\infty$, are, respectively, expressed as
\begin{subequations}
\begin{align}
F_N(q;W)&=\Prb\{X(\varsigma_i)<qY(\varsigma_i), 0\le i<N\},\label{eq:exp-grid}\\
F_\infty(q;W)&=\Prb\{X(\varsigma)<qY(\varsigma), 0\le \varsigma\le W\}.\label{eq:exp-continuum}
\end{align}
\end{subequations}
Continuity of the sample paths implies convergence of the grid maximum to the continuous maximum. %The dimensionless aperture is $\tau=\alpha W$.

\vspace{-2mm}
\subsection{Exact Finite-Port Calculation}
Given (\ref{eq:exp-covariance}), we have the stationary channel recursion
\begin{equation}\label{eq:exp-channel-recursion}
h_{i+1}=\rho h_i+\sqrt{1-\rho^2}\,\xi_i,
\end{equation}
where $\rho=e^{-\Delta}$ according to (\ref{eq:exp-covariance}) and $\xi_i\sim\CN(0,1)$. We have an independent but similar recursion for $g_i$. 

\begin{theorem}[PDF of $f_{\bf X}({\bf x})$]\label{th:pdf-x}
Under the model of (\ref{eq:exp-channel-recursion}), the joint PDF of ${\bf X}=(X_1,\dots,X_N)$ is given by
\begin{equation}\label{eq:exp-gamma-product}
f_{\mathbf X}(\mathbf x)=\sum_{\mathbf k\in\Nzero^{N-1}}\omega_{\mathbf k}\prod_{i=1}^N f_{\mathrm G}(x_i;\kappa_i(\mathbf k)+1,\delta_i),
\end{equation}
where
\begin{equation}\label{eq:wk-exp}
\omega_{\mathbf k}=\frac{c^{|\mathbf k|}}{b^{N-1+2|\mathbf k|}\prod_{j=1}^{N-1}(k_j!)^2}\prod_{i=1}^N\frac{\kappa_i(\mathbf k)!} {\delta_i^{\kappa_i(\mathbf k)+1}},
\end{equation}
and
\begin{equation}
 f_{\mathrm G}(x;\kappa,\delta)=\frac{\delta^{\kappa}}{\Gamma(\kappa)}x^{\kappa-1}e^{-\delta x},
~x>0,
\end{equation}
denotes a Gamma PDF with shape $\kappa>0$ and rate $\delta>0$. In the above, we have defined the index vector ${\bf k}=(k_1,\dots,k_{N-1})$, $k_0=k_N=0$, $\kappa_i({\bf k})=k_{i-1}+k_i$, $c=\rho^2=e^{-2\Delta}$, $b=1-c$, $\delta_1=\delta_N=1/b$ and $\delta_i=\frac{1+c}{b}$ for $i=2,\dots,N-1$. Also, $|\cdot|$ returns the sum of an input vector, i.e., $|\mathbf k|=\sum_{j=1}^{N-1}k_j$.
\end{theorem}

%Let
%\begin{equation}
% f_{\mathrm G}(x;\kappa_{\rm sh},\delta)=\frac{\delta^{\kappa_{\rm sh}}}{\Gamma(\kappa_{\rm sh})}x^{\kappa_{\rm sh}-1}e^{-\delta x},
% \qquad x>0,
%\end{equation}
%denote a Gamma PDF with generic shape $\kappa_{\rm sh}>0$ and rate $\delta>0$. The shape $\kappa_{\rm sh}$ is a placeholder, distinct from the port-dependent powers $\kappa_i(\mathbf k)$ introduced in \eqref{eq:exp-edge-indices}. Regrouping the terms of \eqref{eq:exp-joint-power} then yields
%\begin{equation}
% f_{\mathbf X}(\mathbf x)
% =\sum_{\mathbf k\in\Nzero^{N-1}}\omega_{\mathbf k}
%  \prod_{i=1}^N f_{\mathrm G}(x_i;\kappa_i(\mathbf k)+1,\delta_i).
% \label{eq:exp-gamma-product}
%\end{equation}
%For clarity, the weight before simplification is
%\begin{equation}
% \omega_{\mathbf k}
% =\frac{c^{|\mathbf k|}}
% {b^{N-1+2|\mathbf k|}\prod_{j=1}^{N-1}(k_j!)^2}
% \prod_{i=1}^N\frac{\kappa_i(\mathbf k)!}
% {\delta_i^{\kappa_i(\mathbf k)+1}},
%\end{equation}
%where $|\mathbf k|=\sum_{j=1}^{N-1}k_j$. Since $\sum_i\kappa_i(\mathbf k)=2|\mathbf k|$, substituting the rates $\delta_i$ simplifies this expression to the weights in \eqref{eq:exp-explicit-weights}. They are nonnegative, and integration of \eqref{eq:exp-gamma-product} shows that $\sum_{\mathbf k}\omega_{\mathbf k}=1$.

\begin{proof}
See Appendix \ref{app:proof-pdf-x}.
\end{proof}

%Its power transition density $K_\Delta(x'|x)=f_{X_{i+1}|X_i}(x'|x)$ is given in Appendix~\ref{app:exp-finite-series}. Expanding this density yields the following explicit outage series.
%
%Set $c=\rho^2=e^{-2\Delta}$, and let $\mathbf k=(k_1,\ldots,k_{N-1})\in\Nzero^{N-1}$. With $k_0=k_N=0$, define
%\begin{equation}\label{eq:exp-edge-indices}
%\kappa_i(\mathbf k)\triangleq k_{i-1}+k_i,~\mbox{with }|\mathbf k|=\sum_{j=1}^{N-1}k_j.
%\end{equation}
%The mixture weights are given by
%\begin{equation}\label{eq:exp-explicit-weights}
%\omega_{\mathbf k}=\frac{(1-c)c^{|\mathbf k|}\prod_{i=1}^N \kappa_i(\mathbf k)!}{\prod_{j=1}^{N-1}(k_j!)^2(1+c)^{\sum_{i=2}^{N-1}[\kappa_i(\mathbf k)+1]}}.
%\end{equation}
%For $N=2$, the sum in the exponent is empty and equals zero. The weights are nonnegative and sum to one.

\begin{corollary}[Explicit finite-port series of the outage probability]\label{cor:exp-finite-series}
For $N\ge2$, $q>0$, and $W>0$, we have
\begin{equation}\label{eq:exp-finite-series}
F_N(q;W)=\sum_{\mathbf k,\boldsymbol\ell\in\Nzero^{N-1}}\omega_{\mathbf k}\omega_{\boldsymbol\ell}\prod_{i=1}^NI_p\left(\kappa_i(\mathbf k)+1,\kappa_i(\boldsymbol\ell)+1\right).
\end{equation}
%where $p=q/(1+q)$ and $I_p(a_{\rm B},b_{\rm B})$ is the regularized incomplete beta function with generic positive shape parameters $a_{\rm B},b_{\rm B}$, distinct from the constants $a$ and $b$ used elsewhere,
%\begin{equation}
%\begin{aligned}
%I_p(a_{\rm B},b_{\rm B})&=\frac{1}{B(a_{\rm B},b_{\rm B})}
%\int_0^p u^{a_{\rm B}-1}(1-u)^{b_{\rm B}-1}\,du,\\
%B(a_{\rm B},b_{\rm B})&=\frac{\Gamma(a_{\rm B})\Gamma(b_{\rm B})}{\Gamma(a_{\rm B}+b_{\rm B})}.
%\end{aligned}
%\label{eq:beta:function}
%\end{equation}
All terms are nonnegative and the series converges absolutely.
\end{corollary}

\begin{proof}
See Appendix~\ref{app:exp-finite-series}.
\end{proof}

%Equation~\eqref{eq:exp-finite-series} contains no numerical integrals and its weights require no coefficient recursion. For $N=2$, $\omega_k=(1-c)c^k$. At $q=1$, the two- and three-port cases reduce to
%\begin{subequations}
%\begin{align}
%F_2(1;W)&=\frac12-\frac14\sqrt{1-c},\label{eq:exp-two-check}\\
%F_3(1;W)&=\frac12-\frac14\sqrt{1-c}-\frac18\sqrt{1-c^2},\label{eq:exp-three-check}
%\end{align}
%\end{subequations}
%where $c=e^{-2\Delta}$ is evaluated for the corresponding $N$.

It is worth noting that the result in Corollary \ref{cor:exp-finite-series} is consistent with the general expression in Theorem \ref{thm:main}. Here, the coefficients $\omega_{\bf k}$ are specific to the exponential correlation model.

Before proceeding, it is also useful to note that for a general exponential correlation model, there can be a parameter $\alpha>0$ in (\ref{eq:exp-covariance}) to control the decay rate, i.e., ${\rm COV}(\varsigma,\varsigma')=e^{-\alpha|\varsigma-\varsigma'|}$. In this case, the analysis can be easily extended by absorbing $\alpha$ into the spatial aperture $W$. In other words, the results can be generalized by replacing $W$ by $\alpha W$.

\vspace{-2mm}
\subsection{General $q>0$ in the Infinite-Port Limit}
\begin{theorem}[Infinite-port limit of the outage probability]\label{th:Pout-infty-1}
As $N\to\infty$, the outage probability converges to
\begin{equation}\label{thm-eq:Flimit}
F_\infty(q;W)=\int_{{\cal D}_q}e^{-(x+y)}v(W,x,y)dx dy,
\end{equation}
where $\mathcal D_q=\{(x,y):x,y\geq0, x<qy\}$, and $v(\varsigma,x,y)$ is the solution to the following partial differential equation (PDE)
\begin{equation}\label{thm-eq:v-derivative}
\frac{\partial v}{\partial\varsigma}=2\left[(1-x)\frac{\partial v}{\partial x}+x\frac{\partial^2 v}{\partial x^2}+(1-y)\frac{\partial v}{\partial y}+y\frac{\partial^2 v}{\partial y^2}\right].
\end{equation}
\end{theorem}

\begin{proof}
See Appendix \ref{app:general-q}.
\end{proof}

\begin{corollary}\label{coro:dvdvsigma}
By change of variables, $r=x+y$ and $u=\frac{x}{x+y}$, the asymptotic outage probability can be rewritten as
\begin{equation}\label{thm-eq:Flimit-ru}
F_\infty(q;W)=\int_0^p\int_0^\infty re^{-r}v(W,r,u)dr du,
\end{equation}
where $p=\frac{q}{1+q}$ and $v(\varsigma,r,u)$ is the solution to
\begin{equation}\label{eq:app-pde-ru}
\frac{\partial v}{\partial\varsigma}=2\left[r \frac{\partial^2 v}{\partial r^2}+(2-r)\frac{\partial v}{\partial r}+\frac1r\frac{\partial}{\partial u}\left(u(1-u)\frac{\partial v}{\partial u}\right)\right].
\end{equation}
\end{corollary}

\begin{proof}
With $r$ and $u$, we have $x=ru$ and $y=r(1-u)$, and the outage region becomes $r>0$, $0<u<p$ and its absorbing boundary is $u=p$. The Jacobian of the transformation gives $e^{-(x+y)}dxdy=re^{-r}drdu$, and therefore, we have (\ref{thm-eq:Flimit-ru}).

Next, applying the chain rule gives
\begin{equation}\label{eq:d1}
\frac{\partial v}{\partial x}=\frac{\partial v}{\partial r}+\frac{1-u}{r}\frac{\partial v}{\partial u},~\mbox{and }
\frac{\partial v}{\partial y}=\frac{\partial v}{\partial r}-\frac{u}{r}\frac{\partial v}{\partial u}.
\end{equation}
For the second-order terms, we apply the chain rule once more, and it can be shown that
\begin{equation}\label{eq:d2}
\left\{\begin{aligned}
\frac{\partial^2 v}{\partial x^2}&=\frac{\partial^2 v}{\partial r^2}+\frac{2(1-u)}{r}\frac{\partial^2 v}{\partial r\partial u}+\frac{(1-u)^2}{r^2}\frac{\partial^2 v}{\partial u^2}\\
&\hspace{3mm}-\frac{2(1-u)}{r^2}\frac{\partial v}{\partial u},\\
\frac{\partial^2 v}{\partial y^2}&=\frac{\partial^2 v}{\partial r^2}-\frac{2u}{r}\frac{\partial^2 v}{\partial r\partial u}+\frac{u^2}{r^2}\frac{\partial^2 v}{\partial u^2}+\frac{2u}{r^2}\frac{\partial v}{\partial u}.
\end{aligned}\right.
\end{equation}
Substituting (\ref{eq:d1}) and (\ref{eq:d2}) into (\ref{thm-eq:v-derivative}) gives (\ref{eq:app-pde-ru}).
\end{proof}

\begin{corollary}\label{coro:v-solution}
The solution to (\ref{eq:app-pde-ru}) is given by
%One product in \eqref{eq:app-separated-modes} does not generally equal one at zero length for every $r$ and $u$. To satisfy that initial condition, we combine the product solutions:
\begin{equation}\label{eq:v:r:mu}
v(\varsigma,r,u)=\sum_{j=1}^{\infty}\sum_{k=0}^{\infty}\mathfrak c_{jk}\mathcal V_{jk}(\varsigma,r,u).
\end{equation}
where 
\begin{multline}\label{eq:app-separated-modes}
\mathcal V_{jk}(\varsigma,r,u)=e^{-2(\nu_j+k)\varsigma}\\
\times r^{\nu_j}L_k^{(2\nu_j+1)}(r)P_{\nu_j}(1-2u)
\end{multline}
in which $L_k^{(2\nu_j+1)}(r)$ are defined in \cite[Secs.~18.5 and 18.8]{NIST2010}, $P_\nu(\cdot)$ is the Legendre function \cite[Sec.~14.2(i)]{NIST2010}, and $\nu_1<\nu_2<\cdots$ are the positive solutions to
\begin{equation}\label{eq:exp-roots}
P_{\nu_j}\left(\frac{1-q}{1+q}\right)=0.
\end{equation}
Also, the coefficients $\mathfrak c_{jk}$ are found by
\begin{equation}\label{eq:app-series-coefficients}
\mathfrak c_{jk}=\frac{\mathcal I_j^\phi\mathcal I^{R}_{jk}}{\mathcal I^{\phi^2}_j\mathcal I^{R^2}_{jk}},
\end{equation}
where 
\begin{equation}\label{eq:app-coefficient-integrals}
\left\{\begin{aligned}
\mathcal I^{\phi}_j
&=\int_0^p\phi_j(u)\,du,\\
\mathcal I^{\phi^2}_j
&=\int_0^p\phi_j(u)^2\,du,
\end{aligned}\right.
\left\{\begin{aligned}
\mathcal I^{R}_{jk}
&=\int_0^\infty re^{-r}R_{jk}(r)\,dr,\\
\mathcal I^{R^2}_{jk}
&=\int_0^\infty re^{-r}R_{jk}(r)^2\,dr,
\end{aligned}\right.
\end{equation}
where $\phi_j(u)=P_{\nu_j}(1-2u)$ and $R_{jk}(r)=r^{\nu_j}L_k^{(2\nu_j+1)}(r)$.
\end{corollary}

\begin{proof}
See Appendix \ref{app:proof-v-solution}.
\end{proof}

\begin{theorem}[Asymptotic outage probability as spectral series]\label{th:Pout-infty-2}
The asymptotic outage probability as $N\to\infty$ is given by
\begin{equation}\label{eq:exp-spectral}
F_\infty(q;W)=\sum_{j=1}^{\infty}A_jH_{\nu_j}(\alpha W),
\end{equation}
where
\begin{equation}\label{eq:exp-A-explicit}
%A_j=-\frac{2p(1-p)(2\nu_j+1)}{2\nu_j^2(\nu_j+1)^2}\frac{P'_{\nu_j}(1-2p)}{\dot P_{\nu_j}(1-2p)},
A_j=-\frac{2q}{(1+q)^2}\frac{2\nu_j+1}{\nu_j^2(\nu_j+1)^2}\frac{P'_{\nu_j}\left(\frac{1-q}{1+q}\right)}{\dot P_{\nu_j}\left(\frac{1-q}{1+q}\right)},
\end{equation}
where $P_\nu(z)$ denotes the Legendre function of the first kind of degree $\nu$, evaluated on $(-1,1]$, $P'_\nu(z)=\frac{\partial P_\nu(z)}{\partial z}$, $\dot{P}_{\nu_j}(z)=\left.\frac{\partial P_{\nu}(z)}{\partial\nu}\right|_{\nu=\nu_j}$, and $\{\nu_j\}$ are the positive solutions to (\ref{eq:exp-roots}) and 
%\begin{equation}\label{eq:exp-roots}
%P_{\nu_j}\left(\frac{1-q}{1+q}\right)=0,
%\end{equation}
%and 
\begin{equation}\label{eq:exp-H}
H_\nu(\tau)=\frac{\Gamma(\nu+2)^2}{\Gamma(2\nu+2)}e^{-2\nu\tau} {}_2F_1\left(\nu,\nu;2\nu+2;e^{-2\tau}\right),
\end{equation}
in which ${}_2F_1$ is the Gauss hypergeometric function.
\end{theorem}

\begin{proof}
See Appendix \ref{app:proof-Flimit}.
\end{proof}

Note that in Theorem \ref{th:Pout-infty-2}, the parameter $\alpha$ is reinstated to account for general exponential correlation.

We can have a sanity check for (\ref{eq:exp-spectral}) at zero length, i.e., $W=0$. In this case, \eqref{eq:app-initial-expansion} guarantees that $F_\infty(q;0)=1$, as required.\footnote{To be precise, $F_\infty(q;0)=p=\frac{q}{1+q}$, which corresponds to the outage probability for a fixed antenna receiver subject to interference under Rayleigh fading with the SIR threshold, $q$. However, for any reasonable SIR target $t$, a few users will make $q$ very large, giving $p\approx 1$.} Also, we can see that $H_\nu(0)=1$ from (\ref{eq:exp-H}) because ${}_2F_1\left(\nu,\nu;2\nu+2;1\right)=\frac{\Gamma(2\nu+2)}{\Gamma(\nu+2)^2}$. As such, $\sum_jA_j=p$. On the other hand, the coefficients in \eqref{eq:app-radial} are nonnegative and sum to one at $\tau=0$. Since $k\geq0$, every exponential in that equation is at most $e^{-2\nu\tau}$. Consequently, we have
\begin{equation}\label{eq:app-H-bound}
0\leq H_\nu(\tau)\leq e^{-2\nu\tau}.
\end{equation}
Combining \eqref{eq:app-H-bound} in the above with $A_j\geq0$ and $\sum_jA_j=1$ bounds the terms omitted from summing only up to $J$ terms, giving the following error bound
\begin{align}
0 &\le F_\infty(q;W) -\sum_{j=1}^{J}A_jH_{\nu_j}(\alpha W)\notag\\
&\le \left(1-\sum_{j=1}^{J}A_j\right) e^{-2\nu_{J+1}\alpha W}.\label{eq:exp-tail}
\end{align}

\subsection{Connections to Classical Problems}
The infinite-port analysis connects with several classical problems. First, the condition $P_\nu(1-2p)=0$ that defines the indices $\nu_j$ in \eqref{eq:exp-spectral} is the same eigenvalue condition that governs the survival probability of Brownian motion inside a circular cone, $P_\nu(\cos\theta_0)=0$, where $\theta_0$ is the cone half-angle \cite{Spitzer-1958,Burkholder-1977,DeBlassie-1987,Banuelos-1997}; in our case, $p=\frac{q}{1+q}$ plays the role of the cone angle. Second, the separation of \eqref{eq:app-pde-ru} into a Legendre angular part and a radial part of the form $r^{\nu}L_k^{(2\nu+1)}(r)$ mirrors the separation of the Schr\"odinger equation for the hydrogen atom \cite{Schrodinger-1926}, the only difference being our degrees $\nu_j$ are not integers. Third, the operator $u(1-u)\partial_{uu}+(1-2u)\partial_u$ is, up to a constant factor, the generator of a Wright-Fisher diffusion with symmetric mutation, a classical model for the drift of gene frequencies \cite{Kimura-1955}. Fourth, the power process, the Bessel-$\mathrm I_0$ transition kernel, and the Laguerre expansion coincide with those of the Feller (square-root) diffusion \cite{Feller-1951}, known in finance as the Cox-Ingersoll-Ross process \cite{CIR-1985}. Finally, in the special case $q=1$, the infinite-port outage admits a determinant form that coincides with the Karlin-McGregor formula for coincidence probabilities of non-intersecting paths \cite{Karlin-1959}.

%\subsubsection{Dominant-Mode Approximation}
%For fixed $q$ and increasing
%aperture, its slowest-decaying term gives
%\begin{equation}
%\begin{aligned}
% F_\infty(q;W)
% &\sim A_1
% \frac{\Gamma(\nu_1+2)^2}{\Gamma(2\nu_1+2)}
% e^{-2\nu_1\alpha W},\\
% &\hspace{3cm}\alpha W\to\infty.
%\end{aligned}
%\label{eq:exp-general-large-W}
%\end{equation}
%Here, the infinite-port limit has already been taken
%at fixed $W$. Equation~\eqref{eq:exp-general-large-W}
%then describes the effect of increasing the aperture.
%This differs from both increasing port density at fixed
%aperture and taking the small-threshold limit at fixed
%$N$ in Section~\ref{sec:consequences}.
%These results apply to exponential spatial correlation.
%
%\subsubsection{Outage at a Fixed SIR Target}
%
%An interesting design question is: how does the number
%of users $U$ affect the outage probability for
%a given aperture $W$ and a fixed SIR target $t$?
%
%To answer this question, we substitute the
%physical load into the asymptotic outage expression,
%rather than keeping the normalized threshold fixed:
%\begin{equation}
% P_{\rm out}^{(\infty)}(U,t,W)
% =F_\infty((U-1)t;W).
% \label{eq:fa-physical}
%\end{equation}
%Thus, at fixed $t$ and $W$, increasing $U$
%increases the normalized threshold $q=(U-1)t$,
%and consequently the outage probability.
%In contrast, fixing $q=1$ while varying $U$
%would implicitly reduce the SIR target to
%$t=1/(U-1)$, preventing a fair comparison
%under a common performance requirement.
%Numerical validation is presented in
%Section~\ref{sec:numerics}.

\subsection{Maximum User Load, $U_{\max}$}
With the asymptotic outage probability result in Theorem \ref{th:Pout-infty-2}, it is possible to find the fundamental limit on the maximum number of supportable users, $U_{\max}$, under fast FAMA for a given SIR threshold, $t$, with a prescribed outage probability tolerance, say $\varepsilon\in(0,1)$, for each user with FAS of size $W$. The following theorem presents this fundamental result.

\begin{theorem}\label{thm:Umax}
The maximum number of admissible users, $U_{\max}$, in fast FAMA, subject to the outage probability requirement $F_{\infty}((U-1)t;W)\le\varepsilon$, is given by
\begin{equation}\label{eq:fa-Uexact}
U_{\max}(W,t,\varepsilon)=1+\left\lfloor\frac{q_\varepsilon(W)}{t}\right\rfloor,
\end{equation}
where $\lfloor a\rfloor$ returns the largest integer smaller than $a$, and
\begin{equation}\label{eq:fa-qepsilon-sup}
q_\varepsilon(W)=\sup\{q>0:F_\infty(q;W)\leq\varepsilon\}.
\end{equation}
\end{theorem}

\begin{proof}
The objective is to find the largest integer $U$ satisfying $F_\infty((U-1)t;W)\leq\varepsilon$. Given the SIR target, $t$, fixed, serving more users evidently increases $q=(U-1)t$, and therefore enlarges the outage event. The admissible threshold $q_\varepsilon(W)$ is obtained by solving
\begin{equation}
 F_\infty(q_\varepsilon(W);W)=\varepsilon,
 \label{eq:fa-qepsilon}
\end{equation}
where $F_\infty(q;W)$ is given by the spectral series
in \eqref{eq:exp-spectral}. 

For any $W>0$, continuity and strict monotonicity in $q$ identify $q_\varepsilon(W)$ as the unique solution of $F_\infty(q_\varepsilon(W);W)=\varepsilon$. Substituting $q=(U-1)t$ then gives \eqref{eq:fa-Uexact}. 
\end{proof}

The threshold, $q_\varepsilon$, is generally found by scalar numerical inversion of the spectral series. For a prescribed integer $U$, inversion is unnecessary: evaluate $F_\infty((U-1)t;W)$ directly. Near an integer admission boundary, the series truncation certificate should be used to resolve the comparison with $\varepsilon$.

\vspace{-2mm}
\section{Numerical Results}\label{sec:numerics}
In this section, we present Monte Carlo simulation results to validate our analytical results and utilize those results to understand the performance of fast FAMA systems. Both Jakes and exponential correlation models are considered. Finite-port series evaluations use the coefficient recursion in \eqref{eq:recursion} and the outage expression in \eqref{eq:main}. Monte Carlo simulations generate independent standard proper complex Gaussian vectors and multiply them by a Cholesky factor of $\R$. The maximum of the resulting per-port power ratios is then compared with each threshold. This procedure directly simulates the normalized model in Proposition~\ref{prop:reduction}, with the number of users incorporated through $q=(U-1)t$. Threshold sweeps consider $2\times10^6$ independent realizations per covariance matrix, while the aperture sweep uses $10^6$ realizations per aperture.

\begin{figure}[]
\centering\maybeincludegraphics[width=.9\columnwidth]{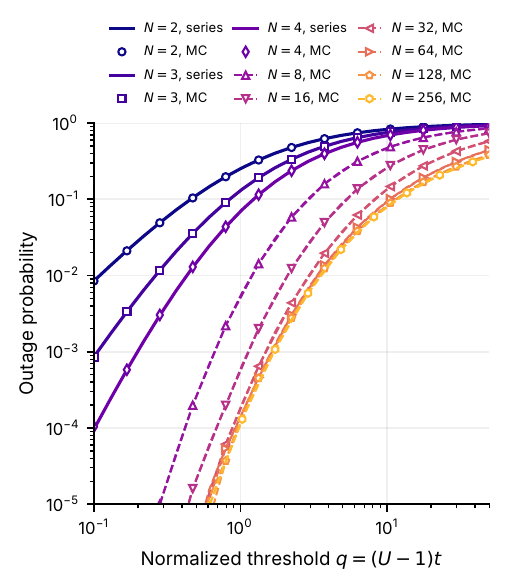}
\caption{General finite-port outage under full Jakes correlation at $W=3$. Solid curves evaluate \eqref{eq:main} with $|\mathbf m|,|\mathbf n|\le24$ for $N=2,3,4$. Open markers use $2\times10^6$ Monte Carlo realizations per covariance matrix; for $N\ge8$, only Monte Carlo results are shown (dashed lines). %The horizontal axis is $q=(U-1)t$, so a physical threshold is obtained by dividing by $U-1$. Legends are placed above the plotting area.
}\label{fig:outage}
\vspace{-3mm}
\end{figure}

%\vspace{-2mm}
%\subsection{General finite-port evaluation}
Fig.~\ref{fig:outage} considers the Jakes model and compares the truncated positive series with Monte Carlo for $N=2,3,4$ at fixed aperture $W=3$, over the range $0.1\le q\le 50$. The total-degree cutoff is $K=24$ in both multi-indices. Solid curves are finite evaluations of the exact series, rather than claims of zero numerical truncation error. The threshold-independent certificate is given by \eqref{eq:certificate}. For $N=2,3,4$, it is approximately $7\times10^{-16}$, $2\times10^{-10}$, and $2\times10^{-7}$ in absolute probability, respectively. Simulation uncertainty and series truncation are distinct sources of numerical uncertainty. In this figure, we also include Monte Carlo results for much denser FAS candidate ports, $N\in\{8,16,32,64,128,256\}$, which are required to obtain more meaningful outage levels at large $q$. For these values of $N$, the positive series is not a practical evaluation tool under Jakes correlation. An aperture of $W=3$ supports only about $2W+1$ significant spatial degrees of freedom, so the smallest eigenvalue of $\R$ collapses as the grid becomes dense; for instance, it is of the order of $10^{-7}$ already at $N=12$. Since the series requires $a\le\lambda_{\min}(\R)$, its convergence rate deteriorates accordingly, while the number of terms grows as $\binom{N+K}{N}^2$. The Monte Carlo curves show that the outage keeps decreasing well beyond $N=2W+1$, because the selection benefits from the sharp SIR peaks that occur near deep fades of the interference, and it saturates only for $N\approx128$--$256$ at this aperture. Even then, the outage at $q=50$ is about $0.36$. With the Monte Carlo simulation results provided, the results indicate that the finite-series is accurate.

%The curves show the behavior over a range of thresholds and retain all Jakes correlations. Because the uniformly spaced port sets for different $N$ are not necessarily nested, these examples should not be interpreted as a proof of monotonicity for every possible grid replacement. Adding ports to an existing set does, however, reduce outage realization by realization.

%\vspace{-2mm}
%\subsection{Exact aperture dependence for three ports}
%Figure~\ref{fig:three} tests \eqref{eq:threejakes} over small and moderate apertures. The formula remains inexpensive even where the general positive series would converge slowly. It approaches $1/2$ as the three ports coalesce, and approaches the independent-port value $1/8$ as $W$ becomes large. Bessel correlations can produce nonmonotone variations as the positions move; the formula retains those variations without a fitted effective-port count.

%\begin{figure}[!t]
%\centering
%\maybeincludegraphics[width=\columnwidth]{figures/fig2_three_port_jakes.pdf}
%\caption{Outage probability versus aperture $W$ for three equally spaced ports under Jakes spatial correlation at $q=1$. The solid line represents the theoretical result in \eqref{eq:threejakes}, and markers denote Monte Carlo simulations. The dotted line represents the independent-port case, for which the outage probability is $1/8$.}\label{fig:three}
%\end{figure}

\begin{figure*}[]
\centering
\maybeincludegraphics[width=.9\textwidth]{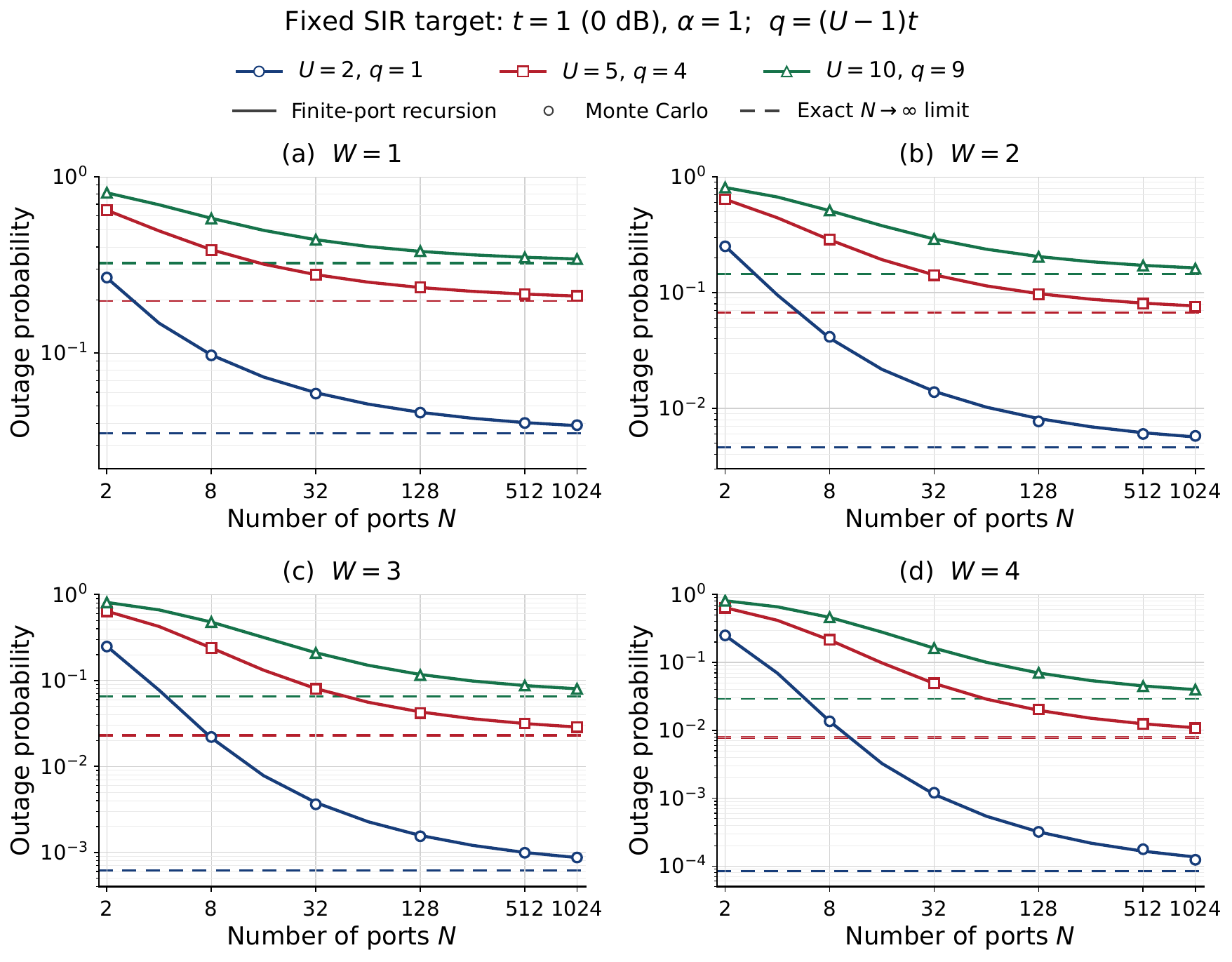}
\caption{Outage probability versus the number of ports $N$ under exponential spatial correlation, with fixed SIR target $t=1$ (0~dB) and $\alpha=1$. Panels (a)--(d) correspond to apertures $W=1,2,3,4$, respectively. The user counts $U=2,5,10$ give $q=(U-1)t=1,4,9$. Solid lines show numerical evaluations of the exact finite-port recursion, and markers denote Monte Carlo simulations. Horizontal dashed lines show $F_\infty(q;W)$ from \eqref{eq:exp-spectral}, the exact continuous-selection outage obtained as $N\to\infty$ at fixed aperture $W$. The horizontal axis uses a base-2 logarithmic scale. The vertical ranges differ between panels.}\label{fig:exp-dense-fast-validation}
\vspace{-3mm}
\end{figure*}

%\vspace{-2mm}
%\subsection{Finite-to-Infinite-Port Comparison at a Fixed SIR Target}
Fig.~\ref{fig:exp-dense-fast-validation} compares the finite-port outage probability with its continuous-selection limit as the number of ports increases at fixed aperture, considering the exponential correlation model. Panels (a)--(d) correspond to $W=1,2,3,4$, respectively, with $\alpha=1$. In every panel, the SIR target is fixed at $t=1$ ($0$~dB), and the user counts $U=2,5,10$ give
\begin{equation}
 q=(U-1)t\in\{1,4,9\}.
\end{equation}
Thus, the figure compares different interference loads under the same physical SIR requirement. The deterministic curves use $N=2^k$, for $k=1,\ldots,10$. Monte Carlo estimates are computed at $N\in\{2,8,32,\allowbreak128,512,1024\}$ by generating both complex channels from their exact stationary Gaussian recursions \eqref{eq:exp-channel-recursion}. Each $(W,N)$ pair uses its own uniform spatial grid. The simulations use $2\times10^5$ independent channel pairs per configuration, increased to $2\times10^6$ for $U=2$, $W\in\{3,4\}$, and $N\in\{128,512,1024\}$ to resolve smaller outage probabilities. Outage is tested only at the $N$ available ports, consistently with the finite-port model. As expected, the solid curves (the finite-series using \eqref{eq:exp-finite-series}) decrease toward the horizontal dashed benchmarks (the asymptotic result using \eqref{eq:exp-spectral}) as the port density increases. Adding users raises both the finite-port outage and its limiting value, whereas increasing the aperture reduces outage for each fixed user count. Moreover, the Monte Carlo markers closely follow the finite-port curves, confirming the accuracy of the finite series. On the other hand, the remaining gap at $N=1024$ is particularly visible for the larger apertures. At $W=4$, the finite-port probabilities are approximately $1.3631\times10^{-4}$, $1.0914\times10^{-2}$, and $3.9755\times10^{-2}$ for $U=2,5,10$, respectively. Their corresponding continuous-selection limits are $8.3873\times10^{-5}$, $7.8784\times10^{-3}$, and $2.9464\times10^{-2}$. Thus, roughly one thousand ports need not eliminate the gap to continuous selection. Moreover, the distinct limiting values show that increasing the port density does not remove the performance penalty of adding users at a fixed target SIR.

\begin{figure*}[]
\centering
\maybeincludegraphics[width=.95\textwidth]{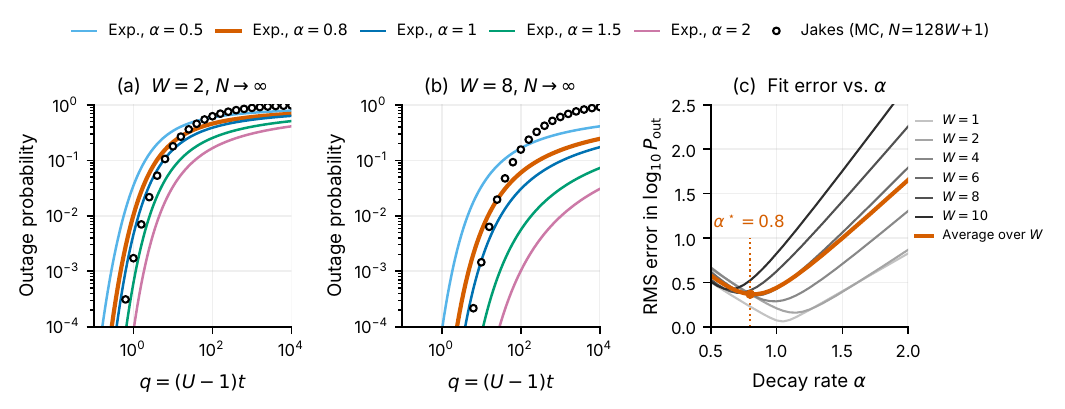}
\caption{Calibration of the exponential decay rate $\alpha$ against Jakes correlation for $N\to\infty$. (a), (b) Outage probability versus $q$ for $W=2$ and $W=8$: lines show $F_\infty(q;W)$ from \eqref{eq:exp-spectral} for several values of $\alpha$, and markers show Jakes outage obtained by Monte Carlo with $N=128W+1$ ports and $10^6$ realizations. (c) RMS error in $\log_{10}$ outage probability versus $\alpha$, evaluated where the Jakes outage lies in $[10^{-3},0.5]$; the thin curves correspond to individual apertures, and the thick curve is their average over the 13 simulated apertures $W\in\{0.5,1,1.5,2,2.5,3,4,\ldots,10\}$, which is minimized at $\alpha^\star=0.8$.}\label{fig:alpha-fit}
\vspace{-3mm}
\end{figure*}

%\vspace{-2mm}
%\subsection{Choice of the Exponential Decay Rate}
The exponential model is usually preferred for its analytical tractability but does not represent a physically grounded model like Jakes. As a result, its decay rate $\alpha$ should be carefully chosen to reproduce the outage behavior of a physically motivated correlation model. The results in Fig.~\ref{fig:alpha-fit} use Jakes correlation as the reference and select $\alpha$ by comparing outage probabilities directly, instead of matching the correlation functions themselves. For each aperture, the Jakes reference is obtained by Monte Carlo simulations with a dense grid of $N=128W+1$ ports, which approximates continuous port selection, and $10^6$ realizations. The exponential counterpart is the exact continuous-selection outage $F_\infty(q;W)$ from \eqref{eq:exp-spectral}. In Figs.~\ref{fig:alpha-fit}(a) and (b), we compare both models for $W=2$ and $W=8$ under several values of $\alpha$. Specifically, a larger $\alpha$ weakens the spatial correlation and lowers the outage probability. Fig.~\ref{fig:alpha-fit}(c) shows the root-mean-square (RMS) error between $\log_{10}$ outage probabilities, computed over the thresholds for which the Jakes outage lies in $[10^{-3},0.5]$. Averaging the squared error over the $13$ simulated apertures $W\in\{0.5,1,1.5,2,2.5,3,4,\ldots,10\}$, with each aperture weighted by its number of valid thresholds, the average RMS error is minimized at $\alpha^\star=0.8$, which will be adopted in the subsequent results. For small apertures, the individual optima are close to $\alpha=1$, while they decrease as $W$ grows. It is important to note that matching the correlation functions in the least-squares sense would instead give $\alpha\approx5$, which badly overestimates the selection gain.

Notably, Fig.~\ref{fig:alpha-fit} also reveals a structural difference between both models. For a fixed aperture, the exponential outage decreases much more rapidly in $q$ than the Jakes outage. This occurs because the exponential correlation corresponds to a non-differentiable channel process, whose rapid spatial fluctuations offer more selection opportunities as the aperture grows. No single value of $\alpha$ can therefore represent Jakes correlation uniformly over large apertures and large thresholds.

\begin{figure}[]
\centering
\maybeincludegraphics[width=.95\columnwidth]{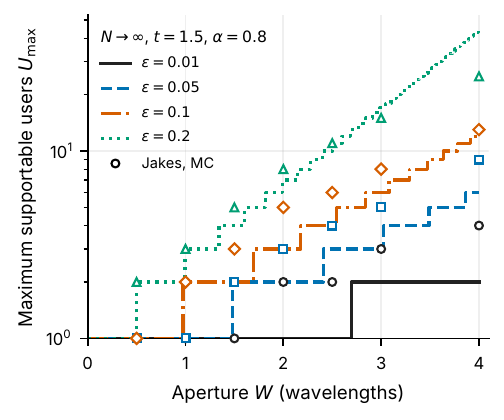}
\caption{Maximum user load versus the FAS aperture under exponential spatial correlation, for $t=1.5$, $\alpha=0.8$, and $\varepsilon\in\{0.01,0.05,0.1,0.2\}$. The curves use the admission rule \eqref{eq:fa-Uexact}; markers show the results under Jakes correlation obtained by Monte Carlo simulations. The vertical axis is logarithmic.}\label{fig:umax-aperture}
\vspace{-3mm}
\end{figure}

\begin{figure}[]
\centering
\maybeincludegraphics[width=.95\columnwidth]{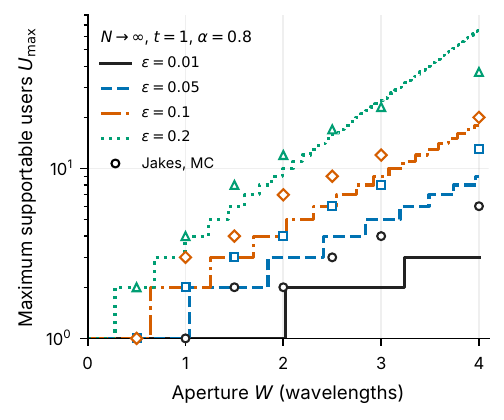}
\caption{Maximum user load versus the FAS aperture under exponential spatial correlation, for $t=1$, $\alpha=0.8$, and $\varepsilon\in\{0.01,0.05,0.1,0.2\}$. The curves use the admission rule \eqref{eq:fa-Uexact}; markers show the result under Jakes correlation obtained by Monte Carlo simulations. The vertical axis is logarithmic.}\label{fig:umax-aperture-t1}
\vspace{-3mm}
\end{figure}

%\vspace{-2mm}
%\subsection{Maximum Supportable Users and Fixed-Target Outage}
Finally, in Figs.~\ref{fig:umax-aperture} and \ref{fig:umax-aperture-t1}, we illustrate the exact maximum user load from \eqref{eq:fa-Uexact}, with $q_\varepsilon(W)$ determined by \eqref{eq:fa-qepsilon}, for FAS apertures up to $W=4$, considering exponential correlation with the calibrated decay rate $\alpha^\star=0.8$, for $t=1.5$ and $t=1$, respectively. Note that the staircases shown are exact: for each integer $U$, the aperture at which $U_{\max}$ jumps from $U-1$ to $U$ is obtained by solving $F_\infty((U-1)t;W)=\varepsilon$ for $W$, which is possible because $F_\infty$ decreases with $W$. As expected, a larger aperture, a relaxed outage requirement, or a lower SIR target supports more users. When $U_{\max}=1$, no additional user is allowed and having multiple users on the same channel use is infeasible at this outage probability requirement $\varepsilon$. The markers in both figures give $U_{\max}$ under the Jakes correlation model, computed from the empirical quantiles of the same dense-grid Monte Carlo samples used in Fig.~\ref{fig:alpha-fit}.

A couple of observations can be made from the figures. First, as the SIR target, $t$, decreases from $1.5$ to $1$, a larger $U_{\max}$ can be achieved, as expected. Similarly, if we move from a more stringent outage probability requirement, $\varepsilon$, to a lesser one, the results again indicate that more users can be accommodated on the same channel use using fast FAMA. On the other hand, we see that the calibrated exponential model seems to track the Jakes results closely, differing by only a few users, when $W\le 3$, although the two models are not supposed to produce the same results. This consistency, nevertheless, shows that the exponential correlation model is not merely a mathematical convenience but is able to capture key channel characteristics of a physically grounded model such as Jakes. That said, for larger apertures, $W>3$, the separation between the models begins to show. In general, for small $\varepsilon$, the exponential model tends to be more conservative than Jakes on $U_{\max}$ while for large $\varepsilon$, the opposite occurs. In particular, for large $W$, the exponential correlation model appears to be overly generous on $U_{\max}$. Thus, the large-aperture results for the exponential model should not be interpreted as predictions for Jakes channels. Regardless, the results confirm that fast FAMA supports massive multiple access, with only receiver-side FAS at the users.

%For instance, at $W=10$, $t=1.5$, and $\varepsilon=0.1$, the exponential model gives $U_{\max}=751$, whereas the Jakes value is $67$. 

\vspace{-2mm}
\section{Conclusion}\label{sec:conclusion}
This paper has established an exact analytical framework for characterizing the outage performance and user scalability of fast FAMA. For an arbitrary finite number of candidate ports and any positive-definite spatial covariance matrix, the outage probability was expressed through a nonnegative Gamma-mixture representation with recursively computable coefficients and a certified truncation bound. Under exponential spatial correlation, the analysis was further extended to the fixed-aperture continuous-selection limit, yielding an explicit spectral representation that enables the fundamental interplay among aperture size, interference load, SIR requirement and outage probability to be quantified. Also, the resulting maximum-user analysis showed that fast FAMA can support a remarkably large number of simultaneously transmitting users when sufficient spatial aperture is available, relying only on receiver-side antenna reconfigurability rather than complex transmitter-side multiuser processing. Numerical results validated the finite-port analysis and continuous-selection limit, while comparisons with Jakes correlation showed that a suitably calibrated exponential model can capture the key performance trends over practically relevant aperture ranges. Overall, the results uncovered the substantial multiple-access potential of fast FAMA and provided a rigorous analytical foundation for understanding its ultimate spatial-interference suppression capability and guiding the design of future scalable FAMA systems.

%This paper derived an exact outage representation for fast FAMA with an arbitrary finite number of ports and a positive-definite spatial covariance matrix. Under constant-modulus modulation, the aggregate complex interference reduces exactly to a single correlated Gaussian vector. The resulting comparison of two independent correlated exponential vectors leads to a nonnegative Gamma-mixture series with computable coefficients and a certified truncation error. Exact special cases and the small-threshold behavior provide additional analytical checks.
%
%For exponential spatial correlation, we also derived the exact continuous-selection limit as $N\to\infty$ at fixed aperture, including a closed form at $q=1$ and a spectral series for general $q$. Comparisons at a fixed SIR target show that increasing the number of users raises both finite-port outage and its limiting value. Increasing port density therefore does not remove the interference penalty or the nonzero outage floor at fixed aperture.
%
%The continuous-selection result further links aperture to the maximum supportable user count. Three design methods are provided: exact numerical inversion of the spectral series, a fully explicit large-load approximation, and an explicit conservative guarantee. The approximation avoids root finding but can lose accuracy at small loads; the guaranteed count provides a sufficient admission condition. All results assume ideal port selection and concern per-user outage. Practical channel estimation and port-selection constraints remain directions for future work.

\appendices
\vspace{-2mm}
%\section{Proofs for Finite-Port Results}\label{app:mixture-details}
\section{Mixture Weights}\label{appA-1}
With $0<a\le\lambda_{\min}(\R)$, all eigenvalues of $\B$ lie in $[0,1)$. We can therefore define a Gaussian vector $\mathbf e\sim\CN(\mathbf0,\B)$, even if $\B$ is singular. For diagonal $\mathbf V$ sufficiently close to zero, the Gaussian quadratic-form identity gives \cite{Mathai-1992}
\begin{equation}
 \frac{1}{\det(\I-\B\mathbf V)}
 =\E\!\left[\exp\!\left(\sum_i \upsilon_i|e_i|^2\right)\right].
\end{equation}
Expanding the exponential and comparing with \eqref{eq:pgf} yields
\begin{equation}\label{eq:wm}
w_{\mathbf m}=\frac{C}{\prod_i m_i!}\E\!\left[\prod_i |e_i|^{2m_i}\right]\ge0.
\end{equation}
Additionally, $\det(\I-\B)=\det(a\R^{-1})=C$. Evaluating the generating function at $\boldsymbol\upsilon=\mathbf1$ shows that $\sum_{\mathbf m}w_{\mathbf m}=1$. %These facts justify the normalized nonnegative mixture in \eqref{eq:gamma} and termwise integration in \eqref{eq:main}.

\vspace{-2mm}
\section{Proof of Theorem \ref{th:pdf-x}}\label{app:proof-pdf-x}
%We prove the explicit finite-port series
%\eqref{eq:exp-finite-series} under exponential correlation.
%For $N\geq2$ and $W>0$, set
%$\Delta=W/(N-1)$, $\rho=e^{-\Delta}$,
%$c=\rho^2$, and $b=1-c$.
%The Gaussian recursion in \eqref{eq:exp-channel-recursion} first gives a conditional power density; expanding its Bessel factor then expresses the joint density as a Gamma mixture. This calculation is for finite $N$. The continuous-port limit is derived separately in Appendix~\ref{app:general-q}.
The model \eqref{eq:exp-channel-recursion} shows that $h_{i+1}\mid h_i$ is complex Gaussian with mean $\rho h_i$ and variance $1-\rho^2$. Its squared magnitude thus has a non-central exponential distribution. Because this distribution only depends on $h_i$ through $|h_i|^2=x$, integrating out the complex phase gives the transition kernel
\begin{align}
K_\Delta(x'\mid x)&=f_{X_{i+1}\mid X_i}(x'\mid x)\notag\\
&=\frac{1}{b}\exp\!\left(-\frac{x'+cx}{b}\right)\mathrm I_0\!\left(\frac{2\sqrt{cxx'}}{b}\right),\label{eq:exp-kernel}
\end{align}
in which $\mathrm I_0$ is the modified Bessel function of the first kind of order zero (which is not to be confused by the regularized incomplete beta function $I_p$). The interference powers have the same kernel, and independence between the two channels, $X$ and $Y$, gives the conditional joint density
\begin{equation}
f_{X_{i+1},Y_{i+1}\mid X_i,Y_i}(x',y'\mid x,y)=K_\Delta(x'\mid x)K_\Delta(y'\mid y).
\end{equation}

For $N\geq2$, by the Markov property, the joint density of the $N$ port powers factorizes as
\begin{equation}\label{eq:exp-joint-markov}
f_{\mathbf X}({\bf x})=e^{-x_1}\prod_{i=1}^{N-1} K_\Delta(x_{i+1}\mid x_i).
\end{equation}
Substituting \eqref{eq:exp-kernel} into the above gives
\begin{multline}\label{eq:exp-joint-power}
f_{\mathbf X}(\mathbf x)=b^{-(N-1)}\exp\!\left(-\sum_{i=1}^N \delta_i x_i\right)\\
\times\prod_{i=1}^{N-1}\mathrm I_0\!\left(\frac{2\sqrt{c x_i x_{i+1}}}{b}\right).
\end{multline}
The appearance of a Bessel function in each neighboring-port factor follows from the conditional power density. Expanding the Bessel factor into a nonnegative power series, $\mathrm I_0\!\left(\frac{2\sqrt{cxy}}{b}\right)=\sum_{k=0}^{\infty}\frac{c^k x^k y^k}{b^{2k}(k!)^2}$, (\ref{eq:exp-joint-power}) can be reexpressed as
\begin{equation}\label{eq:exp-joint-power-2}
f_{\mathbf X}(\mathbf x)=b^{-(N-1)}e^{-\sum_{i=1}^N \delta_i x_i}\prod_{i=1}^{N-1}\sum_{k=0}^{\infty}\frac{c^k x_i^k x_{i+1}^k}{b^{2k}(k!)^2}.
\end{equation}
Using (\ref{eq:useful-1}) (see top of next page) and rearranging terms, we further rewrite $f_{\mathbf X}(\mathbf x)$ as
\begin{figure*}
\begin{align}
\prod_{i=1}^{N-1}\sum_{k=0}^{\infty}\frac{c^k x_i^k x_{i+1}^k}{b^{2k}(k!)^2}&=\left(\sum_{k_1=0}^{\infty}\frac{c^{k_1} x_1^{k_1} x_2^{k_1}}{b^{2k_1}(k_1!)^2}\right)\left(\sum_{k_2=0}^{\infty}\frac{c^{k_2} x_2^{k_2} x_3^{k_2}}{b^{2k_2}(k_2!)^2}\right)\cdots \left(\sum_{k_{N-1}=0}^{\infty}\frac{c^{k_{N-1}} x_{N-1}^{k_{N-1}} x_N^{k_{N-1}}}{b^{2k_{N-1}}(k_{N-1}!)^2}\right)\notag\\
&=\sum_{k_1,\dots,k_{N-1}\ge 0}\frac{c^{k_1+\cdots+k_{N-1}}}{b^{2(k_1+\cdots+k_{N-1})}(k_1!)^2\cdots(k_{N-1}!)^2}x_1^{k_1}x_2^{k_1+k_2}\cdots x_N^{k_{N-1}}=\sum_{{\bf k}}\frac{c^{|{\bf k}|}}{b^{2|{\bf k}|\prod_{j=1}^{N-1}(k_j!)^2}}\prod_{i=1}^Nx_i^{\kappa_i({\bf k})}\label{eq:useful-1}
\end{align}
\hrulefill
\end{figure*}
\begin{equation}
f_{\mathbf X}(\mathbf x)=\sum_{{\bf k}\in\mathbb{N}_0^{N-1}}\frac{c^{|{\bf k}|}}{b^{N-1+2|{\bf k}|}\prod_{j=1}^{N-1}(k_j!)^2}\prod_{i=1}^Nx_i^{\kappa_i({\bf k})}e^{-\delta_ix_i}.
\end{equation}
Recognizing the Gamma PDF, $f_{\rm G}(x_i;\kappa_i({\bf k})+1,\delta_i)$, inside each product, $\omega_{\bf k}$ in (\ref{eq:wk-exp}) can be identified, and we then obtain (\ref{eq:exp-gamma-product}), which completes the proof. Note that all $\omega_{\bf k}$ are nonnegative, and integration of \eqref{eq:exp-gamma-product} confirms that $\sum_{\mathbf k}\omega_{\mathbf k}=1$.

\vspace{-2mm}
\section{Proof of Corollary \ref{cor:exp-finite-series}}\label{app:exp-finite-series}
The proof follows similar steps as in the proof of Theorem \ref{thm:main}. To start with, we note that ${\bf X}$ and ${\bf Y}$ are independent and have equal Gamma rates at each port, see (\ref{eq:exp-gamma-product}). Consequently, the outage probability is given by
\begin{equation}\label{eq:FN}
F_N(q;W)=\sum_{{\bf k},\boldsymbol{\ell}}\omega_{\bf k}\omega_{\boldsymbol{\ell}}\prod_{i=1}^N\mathbb{P}(X_i<qY_i\mid\mathbf k,\boldsymbol\ell).
\end{equation}
The comparison between two Gamma random variables at each port gives the probability in the form of a regularized incomplete beta function.  That is,
\begin{equation}\label{eq:xycompare}
\Prb(X_i<qY_i\mid\mathbf k,\boldsymbol\ell)=I_p\big(\kappa_i(\mathbf k)+1,\kappa_i(\boldsymbol\ell)+1\big).
\end{equation}
Substituting (\ref{eq:xycompare}) into (\ref{eq:FN}) gives the desired result.

%Multiplying these factors across ports and averaging over the two sets of mixture weights proves \eqref{eq:exp-finite-series}. Since each incomplete-beta factor lies in $[0,1]$, omission of mixture-index pairs outside the chosen truncation region is at most the omitted weight, proving \eqref{eq:exp-series-certificate}. Thus, the finite-port series and the integral recursion describe the same event, but the series avoids numerical quadrature.

\vspace{-2mm}
\section{Proof of Theorem \ref{th:Pout-infty-1}}\label{app:general-q}
%We derive the outage probability by starting from its original $2N$-dimensional integral. Under exponential correlation, the Markov structure allows the integrations over the future ports to be grouped into a conditional probability $v_k(x,y)$. This function satisfies an integral recursion, whereas the initial powers are averaged separately with density $e^{-x-y}$. In the dense-port limit, it is the recursion for $v_k$, not this final averaging integral, that leads to a backward Kolmogorov partial differential equation (PDE). We solve that PDE and then average its solution over the initial powers to obtain the infinite-port outage probability. The separate Gamma-mixture derivation for finite $N$ is given in Appendix~\ref{app:exp-finite-series}.
%
%Throughout, $X_i=|h_i|^2$ and $Y_i=|g_i|^2$ denote the desired-signal and normalized-interference powers introduced in Proposition~\ref{prop:reduction}, and $q=(U-1)t>0$. We first use a unit spatial decay rate. For covariance $e^{-\alpha|\varsigma-\varsigma'|}$, the continuum results follow by evaluating the unit-rate solution at the normalized length $\tau=\alpha W$.
To begin with, the following lemmas are useful.

\begin{lemma}\label{lemma:f-taylor}
For any smooth function $f(x,y)$, with $x'=x+\Delta X$ and $y'=y+\Delta Y$ for some small $\Delta X$ and $\Delta Y$, $f(x',y')$ admits the following expansion
\begin{multline}
f(x',y')=f+f_x\Delta X+f_y\Delta Y\\
+\frac{1}{2}\left(f_{xx}\Delta X^2+2f_{xy}\Delta X\Delta Y+f_{yy}\Delta Y^2\right)+\cdots,
\end{multline}
where $f_x=\frac{\partial f}{\partial x}$, $f_y=\frac{\partial f}{\partial y}$, $f_{xx}=\frac{\partial^2 f}{\partial x^2}$, $f_{xy}=\frac{\partial^2 f}{\partial y\partial x}$, $f_{yy}=\frac{\partial^2 f}{\partial y^2}$.
\end{lemma}

\begin{proof}
This directly comes from Taylor series expansion.
\end{proof}

\begin{lemma}\label{lemma:e-result}
Given the model (\ref{eq:exp-channel-recursion}) and an identical model for $g_{i+1}$, we write $X_i=x$, $X_{i+1}=x'=x+\Delta X$, $Y_i=y$ and $Y_{i+1}=y'=y+\Delta Y$. Then we have
\begin{equation}\label{eq:lemma2}
\left\{\begin{aligned}
\mathbb{E}[\Delta X\mid x]&=2(1-x)\Delta+O(\Delta^2),\\
\mathbb{E}[\Delta Y\mid y]&=2(1-y)\Delta+O(\Delta^2),\\
{\rm VAR}[\Delta X\mid x]&=2(1-2\Delta)x+O(\Delta^2),\\
{\rm VAR}[\Delta Y\mid y]&=2(1-2\Delta)y+O(\Delta^2),\\
\mathbb{E}[\Delta X^2\mid x]&=4x\Delta+O(\Delta^2),\\
\mathbb{E}[\Delta Y^2\mid y]&=4y\Delta+O(\Delta^2),\\
\mathbb{E}[\Delta X\Delta Y\mid x,y]&=O(\Delta^2).
\end{aligned}\right.
\end{equation}
\end{lemma}

\begin{proof}
According to (\ref{eq:exp-channel-recursion}), we have
\begin{equation}\label{eq:design-power-step}
X'\mid X=x\ \triangleq\left|\rho\sqrt{x}+\sqrt{b}\,\xi\right|^2,~\xi\sim\CN(0,1),
\end{equation}
where $b=1-\rho^2=1-e^{-2\Delta}=2\Delta+O(\Delta^2)$. Thus,
\begin{equation}\label{eq:app-power-increment}
\Delta X=b(|\xi|^2-x)+2\rho\sqrt{bx}\,\Re\xi.
\end{equation}
Also, we have $\Delta Y$ in a similar form. From (\ref{eq:app-power-increment}) and knowing that $X$ and $Y$ are independent, (\ref{eq:lemma2}) can be obtained.
\end{proof}

\begin{lemma}\label{lemma:dG}
With the conditional PDFs, $K_\Delta(x'\mid x)$ and $K_\Delta(y'\mid y)$ and a smooth function $f(x',y')$, we have
\begin{multline}\label{eq:int-Dq}
\int_{{\cal D}_q} K_\Delta(x'\mid x)K_\Delta(y'\mid y)f(x',y') dx' dy'\\
=f+\Delta{\cal G}f+O(\Delta^2),
\end{multline}
where ${\cal G}f=2[(1-x)f_x+xf_{xx}+(1-y)f_y+yf_{yy}]$.
\end{lemma}

\begin{proof}
Starting with the left-hand side of (\ref{eq:int-Dq}), we can relax the region for the integration from ${\cal D}_q$ to the full $(\mathbb{R}^+)^2$ because an extremely small $\Delta$ is under consideration. Then we expand $f(x',y')$ using Lemma \ref{lemma:f-taylor} and utilize the results in Lemma \ref{lemma:e-result} to obtain (\ref{eq:int-Dq}), which completes the proof.
\end{proof}

\begin{lemma}\label{lemma:v-derivative}
Defining the recursive relationship based on the transition kernel (\ref{eq:exp-joint-markov}) and the one for ${\bf Y}$, 
\begin{multline}\label{lemma-eq:vk}
v_k(x_1,y_1)\triangleq\int_{{\cal D}_q}K_{\Delta}(x_{k+1}\mid x_k)K_{\Delta}(y_{k+1}\mid y_k)\\
\times v_{k-1}(x_1,y_1)dx_{k+1}dy_{k+1},
\end{multline}
for $k=1,\dots,N-1$, with $v_0(x,y)\triangleq 1$, and then writing
\begin{equation}\label{lemma-eq:v}
\left\{\begin{aligned}
v^{(\Delta)}(\varsigma,x,y)&=v_{k-1}(x,y),\\
v^{(\Delta)}(\varsigma+\Delta,x,y)&=v_{k}(x,y),
\end{aligned}\right.
\end{equation}
where $\varsigma$ specifies the covered spatial aperture for FAS in the function $v(\cdot,x,y)$, as $\Delta\to 0$, (\ref{thm-eq:v-derivative}) is true.
\end{lemma}

\begin{proof}
By definitions in (\ref{lemma-eq:vk}) and (\ref{lemma-eq:v}), we have, for small $\Delta$,
\begin{align}
&\frac{v^{(\Delta)}(\varsigma+\Delta,x,y)-v^{(\Delta)}(\varsigma,x,y)}{\Delta}\notag\\
&\,=\frac{1}{\Delta}\left[\begin{array}{r}
\int_{{\cal D}_q}K_{\Delta}(x'\mid x)K_{\Delta}(y'\mid y)v^{(\Delta)}(\varsigma,x',y')dx'dy'\\
-v^{(\Delta)}(\varsigma,x,y)
\end{array}\right]\notag\\
&\stackrel{(a)}{\approx}\frac{1}{\Delta}\left[v^{(\Delta)}(\varsigma,x,y)+\Delta{\cal G}v^{(\Delta)}(\varsigma,x,y)-v^{(\Delta)}(\varsigma,x,y)\right]\notag\\
&\,={\cal G}v^{(\Delta)}(\varsigma,x,y)\notag\\
&\stackrel{(b)}{=}2\left[(1-x)v_x^{(\Delta)}+xv^{(\Delta)}_{xx}+(1-y)v^{(\Delta)}_y+yv^{(\Delta)}_{yy}\right],
\end{align}
where $(a)$ and $(b)$ come from Lemma \ref{lemma:dG}. Finally, as $\Delta\to 0$, we get (\ref{thm-eq:v-derivative}), which completes the proof.
\end{proof}

Now, we are ready to prove Theorem \ref{th:Pout-infty-1}. Given (\ref{eq:exp-joint-markov}) for both ${\bf X}$ and ${\bf Y}$ and that ${\bf X}$ and ${\bf Y}$ are independent, we can write
\begin{multline}
F_N(q;W)=\int_{{\cal D}_q}e^{-(x_1+y_1)}\\
\times\left[\int_{{\cal D}_q}\cdots\begin{array}{c}
K_{\Delta}(x_N\mid x_{N-1})K_{\Delta}(y_N\mid y_{N-1})\\
K_{\Delta}(x_{N-1}\mid x_{N-2})K_{\Delta}(y_{N-1}\mid y_{N-2})\\
\vdots\\
K_{\Delta}(x_2\mid x_1)K_{\Delta}(y_2\mid y_1)\\
\prod_{j=2}^Ndx_jdy_j
\end{array}\right]\\
\times dx_1dy_1,
\end{multline}
which according to (\ref{lemma-eq:vk}) in Lemma \ref{lemma:v-derivative}, gives
\begin{equation}
F_N(q;W)=\int_{{\cal D}_q}e^{-(x_1+y_1)}v_{N-1}(x_1,y_1)dx_1dy_1.
\end{equation}
As $N\to\infty$, we have arrived (\ref{thm-eq:Flimit}). Finally from Lemma \ref{lemma:v-derivative}, we know that $v(\varsigma,x,y)$ is the solution to the PDE (\ref{thm-eq:v-derivative}).

\section{Proof of Corollary \ref{coro:v-solution}}\label{app:proof-v-solution}
We start the proof by presenting the following lemmas which will be important to construct the solution.

\begin{lemma}[Legendre equation]\label{lemma:legend}
The Legendre equation
\begin{equation}\label{eq:app-legendre-equation}
(1-\chi^2)\psi''(\chi)-2\chi\psi'(\chi)+\nu(\nu+1)\psi(\chi)=0,
\end{equation}
has the following solution $\psi(\chi)=C_PP_\nu(\chi)+C_QQ_\nu(\chi)$, where $P_\nu$ and $Q_\nu$ are the two Legendre functions. If $\psi(1)$ is finite, then $C_Q=0$ and by setting $C_P=1$,
\begin{equation}
\psi(\chi)=P_\nu(\chi).
\end{equation}
\end{lemma}

\begin{proof}
See \cite[Sec.~14.2]{NIST2010}.
\end{proof}

\begin{lemma}\label{lemma:R-PDE}
The solution to the following PDE
\begin{equation}\label{eq:app-radial-nu}
rR''+(2-r)R'-\frac{\nu(\nu+1)}{r}R+\mu R=0,
\end{equation}
is given by
\begin{equation}\label{eq:app-R-solution}
R_{k}(r)=r^{\nu}L_k^{(2\nu+1)}(r),~k=0,1,2,\dots,
\end{equation}
where $L_k^{(2\nu+1)}(r)$ are given in \cite[Secs.~18.5 and 18.8]{NIST2010}.
\end{lemma}

\begin{proof}
To find the solution, we first need to remove the term containing $R/r$ so that (\ref{eq:app-radial-nu}) becomes an equation with known solutions. Trying $R(r)\propto r^a$ in the most singular terms, i.e., $r^{a-1}$, of \eqref{eq:app-radial-nu} near $r=0$, requires that $a(a+1)=\nu(\nu+1)$ so that the singular terms vanish. This gives two possibilities $a=\nu$ and $a=-\nu-1$ but the second case diverges at zero. Therefore, the solution to (\ref{eq:app-radial-nu}) takes the form
\begin{equation}\label{eq:app-radial-substitution}
R(r)=r^{\nu}G(r).
\end{equation}
Substituting \eqref{eq:app-radial-substitution} into \eqref{eq:app-radial-nu}, then dividing by $r^{\nu}$, and collecting the remaining terms give
\begin{equation}\label{eq:app-G-equation}
rG''+(2\nu+2-r)G'+(\mu-\nu)G=0,
\end{equation}
which has the generalized Laguerre form $rG''+(\eta_{\rm L}+1-r)G'+kG=0$, with $\eta_{\rm L}=2\nu+1$ and $k=\mu-\nu$. For $k=0,1,2,\ldots$, its solutions include the polynomials $L_k^{(2\nu_j+1)}(r)$ \cite[Secs.~18.5 and 18.8]{NIST2010}, i.e., $G(r)=L_k^{(2\nu+1)}(r)$, for $k=0,1,2,\dots$. These are the solutions selected by requiring regular behavior at zero and a finite integral $\int_0^\infty re^{-r}R(r)^2\,dr$. The solution regular at zero grows too rapidly at infinity unless the series for $G$ terminates as a polynomial. This requires $\mu-\nu$ to be a nonnegative integer. Using this in (\ref{eq:app-radial-substitution}) gives (\ref{eq:app-R-solution}).
%Consequently,
%\begin{equation}
%G(r)=L_k^{(2\nu_j+1)}(r),
%\qquad
%\mu=\nu_j+k,\quad k=0,1,2,\ldots.
%\label{eq:app-G-solution}
%\end{equation}
%Here $j$ selects a solution for the power fraction, and $k$ is the degree of the polynomial used for the total power.
\end{proof}

To solve (\ref{eq:app-pde-ru}), we first note that $v(\varsigma,r,u)$ is the conditional outage probability given $r$ (the initial total channel power) and $u$ (the initial desired-power fraction). If $\varsigma=0$ meaning the FAS has no physical size, outage must occur. Also, the outage boundary happens where $u=p=\frac{q}{1+q}$, implying that $v(\varsigma,r,p)=0$. Hence, we have the initial conditions:
\begin{equation}\label{eq:design-backward}
\left\{\begin{aligned}
v(0,r,u)=1, &~r,u>0,\ u<p,\\
v(\varsigma,r,p)=0, &~\varsigma,r>0.
\end{aligned}\right.
\end{equation}

To construct the solution, we seek solutions of the form that are products of three functions:
\begin{equation}\label{eq:app-product-solution}
\mathcal V(\varsigma,r,u)=T(\varsigma)R(r)\phi(u).
\end{equation}
The functions $T$, $R$, and $\phi$ describe the dependence on $\varsigma$, $r$, and $u$, respectively. This construction is not only motivated by facilitation of finding the solution to (\ref{eq:app-pde-ru}) but also comes from the intuition that the three parameters, $\varsigma$, $r$, and $u$, should play independent roles in affecting the outage probability. 

Substituting \eqref{eq:app-product-solution} into \eqref{eq:app-pde-ru} gives $T'R\phi=2T[rR''\phi+(2-r)R'\phi+(R/r)[u(1-u)\phi']']$. Dividing by $2TR\phi$, wherever the product is nonzero, yields
\begin{equation}\label{eq:app-separated-balance}
\frac{T'(\varsigma)}{2T(\varsigma)}=\frac{rR''(r)+(2-r)R'(r)}{R(r)}+\frac{[u(1-u)\phi'(u)]'}{r\phi(u)}.
\end{equation}
The left side depends only on $\varsigma$, whereas the right side depends only on $r$ and $u$. Varying $\varsigma$ while keeping $r$ and $u$ fixed shows that both sides must equal a constant, say $-\mu$.

The left side of \eqref{eq:app-separated-balance} therefore gives
\begin{equation}\label{eq:app-time-separation}
T'(\varsigma)+2\mu T(\varsigma)=0,
\end{equation}
which has the standard solution
\begin{equation}\label{eq:app-T-solution}
T(\varsigma)=C_Te^{-2\mu\varsigma},
\end{equation}
for some constant $C_T$, which will be addressed subsequently in the final form of the overall solution.

Using the same trick, we can also separate $r$ from $u$. Setting the right side of \eqref{eq:app-separated-balance} equal to $-\mu$, multiplying by $r$ on both sides, and moving the terms involving $R$ to the right gives
\begin{equation}\label{eq:app-r-u-separation}
\frac{[u(1-u)\phi'(u)]'}{\phi(u)}=-\mu r -\frac{r[rR''(r)+(2-r)R'(r)]}{R(r)}.
\end{equation}
The left side depends only on $u$, and the right side only on $r$. Both must therefore equal another constant, say $-\Lambda$.

Multiplying the left side of \eqref{eq:app-r-u-separation} by $\phi(u)$ gives
\begin{equation}\label{eq:design-angular-separated}
-\frac{d}{du}\left[u(1-u)\phi'(u)\right]=\Lambda\phi(u),
\end{equation}
while setting the right side equal to $-\Lambda$ and multiplying by $R(r)/r$ yields another differential equation
\begin{equation}\label{eq:design-radial}
rR''(r)+(2-r)R'(r)-\frac{\Lambda}{r}R(r)+\mu R(r)=0.
\end{equation}

We first solve \eqref{eq:design-angular-separated} for $\phi$ and determine the possible values of $\Lambda$. To do so, we expand its derivative gives
\begin{equation}\label{eq:dDu}
u(1-u)\phi''(u)+(1-2u)\phi'(u)+\Lambda\phi(u)=0.
\end{equation}
By the change of variable, $\chi=1-2u$ and $\psi(\chi)=\phi\left(\frac{1-\chi}{2}\right)$. We have $u(1-u)=(1-\chi^2)/4$, $d\chi/du=-2$, $\phi'(u)=-2\psi'(\chi)$ and $\phi''(u)=4\psi''(\chi)$. If we also write $\Lambda=\nu(\nu+1)$, (\ref{eq:dDu}) can be rewritten as the Legendre equation (\ref{eq:app-legendre-equation}). Therefore, according to Lemma \ref{lemma:legend}, the solution takes the form
\begin{equation}\label{eq:app-phi-solution}
\phi(u)=P_\nu(1-2u).
\end{equation}
This expression has used the normalization $C_P=1$ in Lemma \ref{lemma:legend}, which ensures that the outage probability is one when $u=0$, i.e., $\phi(0)=P_\nu(1)=1$. Additionally, it remains to impose $\phi(p)=0$, i.e., the outage probability is zero when $u=p$. As a result, substituting $u=p$ into \eqref{eq:app-phi-solution} yields (\ref{eq:exp-roots}), giving the positive solutions $\nu_1<\nu_2<\cdots$. Hence,
\begin{equation}\label{eq:app-phi-family}
\phi_j(u)=P_{\nu_j}(1-2u),~\mbox{with }\Lambda_j=\nu_j(\nu_j+1).
\end{equation}
The index $j$ identifies the different solutions that satisfy the condition at $u=p$. We next use each $\Lambda_j$ from the solution of (\ref{eq:exp-roots}) to solve the equation for the total power, $r$.

Substituting $\Lambda=\nu_j(\nu_j+1)$ from \eqref{eq:app-phi-family} into \eqref{eq:design-radial} gives (\ref{eq:app-radial-nu}). According to Lemma \ref{lemma:R-PDE}, we have the solution (\ref{eq:app-R-solution}) so that
\begin{equation}\label{eq:app-R-solution-2}
R_{jk}(r)=r^{\nu_j}L_k^{(2\nu_j+1)}(r),~k=\mu-\nu_j=0,1,2,\dots.
\end{equation}
From the above, it is known that $\mu=\nu_j+k$ and using this substitution into \eqref{eq:app-T-solution} gives the solution for $T(\varsigma)$ as
\begin{equation}\label{eq:app-T-jk}
T_{jk}(\varsigma)=e^{-2(\nu_j+k)\varsigma}.
\end{equation}
Collecting all the solutions (\ref{eq:app-phi-solution}), (\ref{eq:app-R-solution-2}) and (\ref{eq:app-T-jk}) and using them in (\ref{eq:app-product-solution}) yield ${\cal V}_{jk}(\varsigma,r,u)$ in (\ref{eq:app-separated-modes}). The final solution for $v(\varsigma,r,u)$ is then formed by linearly combining all the ${\cal V}_{jk}(\varsigma,r,u)$, resulting in (\ref{eq:v:r:mu}). This construction is possible because the PDE is linear: sums of solutions give another solution, provided the series converges appropriately. Note that $C_T$ in (\ref{eq:app-T-solution}) is conveniently absorbed in the coefficients $\mathfrak c_{jk}$.

The remaining task is to determine the coefficients $\mathfrak c_{jk}$. The idea is to choose them so that the conditions in \eqref{eq:design-backward} are met.

Setting $\varsigma=0$ in \eqref{eq:app-separated-modes} and \eqref{eq:v:r:mu} gives
\begin{equation}\label{eq:app-initial-expansion}
1=\sum_{j=1}^{\infty}\sum_{k=0}^{\infty}\mathfrak c_{jk}R_{jk}(r)\phi_j(u).
\end{equation}
To isolate a particular coefficient for evaluation, we use the fact that $\int_0^p\phi_j(u)\phi_m(u)\,du=0$ for $j\ne m$, which comes from the orthogonality of the Legendre functions. Similarly, for a fixed $j$, the functions $R_{jk}$ have the corresponding property $\int_0^\infty re^{-r}R_{jk}(r)R_{j\ell}(r)\,dr=0$ for $k\ne\ell$.\footnote{The weight, $re^{-r}$, comes from the transformation $x=ru$, $y=r(1-u)$, and $dx\,dy=r\,dr\,du$, so $e^{-(x+y)}\,dx\,dy=re^{-r}\,dr\,du$.}

%For compact notation, define the four integrals
%\begin{equation}
%\begin{aligned}
%\mathcal J_j^{(u)}
%&=\int_0^p\phi_j(u)\,du,\\
%\mathcal N_j^{(u)}
%&=\int_0^p\phi_j(u)^2\,du,\\
%\mathcal J_{jk}^{(r)}
%&=\int_0^\infty re^{-r}R_{jk}(r)\,dr,\\
%\mathcal N_{jk}^{(r)}
%&=\int_0^\infty re^{-r}R_{jk}(r)^2\,dr.
%\end{aligned}
%\label{eq:app-coefficient-integrals}
%\end{equation}
%These symbols are only abbreviations for the integrals needed to calculate the coefficients.

Multiply \eqref{eq:app-initial-expansion} by $re^{-r}R_{jk}(r)\phi_j(u)$ and integrate over $r>0$ and $0<u<p$. All terms with different indices integrate to zero, leaving $\mathcal I^{\phi}_j\mathcal I_{jk}^{R}=\mathfrak c_{jk}\mathcal I_j^{\phi^2}\mathcal I_{jk}^{R^2}$. Consequently, we have obtained (\ref{eq:app-series-coefficients}), which completes the proof.

\section{Proof of Theorem \ref{th:Pout-infty-2}}\label{app:proof-Flimit}
Substituting (\ref{eq:v:r:mu}) into (\ref{thm-eq:Flimit-ru}), expressing $\mathfrak c_{jk}$ using (\ref{eq:app-series-coefficients}) and expressing the integration using the definitions (\ref{eq:app-coefficient-integrals}) give
\begin{equation}\label{eq:app-outage-squared-integrals}
F_\infty(q;W)=\sum_{j=1}^{\infty}\sum_{k=0}^{\infty}
\underbrace{\frac{\left[\mathcal I_j^{\phi}\right]^2}{\mathcal I_j^{\phi^2}}}_{A_j}
\underbrace{\frac{\left[\mathcal I_{jk}^{R}\right]^2}{\mathcal I_{jk}^{R^2}}}_{B_{jk}}
e^{-2(\nu_j+k)\alpha W},
\end{equation}
where $\alpha$ denotes the exponential correlation parameter.

To evaluate $A_j$ which is given by
\begin{equation}\label{eq:app-A-from-integrals}
A_j\triangleq\frac{\left[\mathcal I_j^{\phi}\right]^2}{\mathcal I_j^{\phi^2}}=\frac{\left[\int_0^p\phi_j(u)\,du\right]^2}{\int_0^p\phi_j(u)^2\,du},
\end{equation}
we integrate (\ref{eq:design-angular-separated}) once to obtain
\begin{equation}\label{eq:int-phi}
\int_0^p\phi_j(u)\,du=-\frac{p(1-p)\phi_j'(p)}{\nu_j(\nu_j+1)}.
\end{equation}
Then for the denominator of (\ref{eq:app-A-from-integrals}), we differentiate (\ref{eq:design-angular-separated}) with respect to $\nu$, multiply by $\phi_j$, and subtract the original equation multiplied by $\partial_\nu P_\nu(1-2u)|_{\nu=\nu_j}$. Then integrating $\phi^2$ over the outage region gives
\begin{equation}\label{eq:int-phi2}
\int_0^p\phi_j(u)^2\,du=\frac{p(1-p)\phi_j'(p)\dot P_{\nu_j}(1-2p)}{2\nu_j+1}.
\end{equation}
Substituting (\ref{eq:int-phi}) and (\ref{eq:int-phi2}) into (\ref{eq:app-A-from-integrals}), using the fact that $\phi_j'(p)=-2P'_{\nu_j}(1-2p)$ and noting $p=\frac{q}{1+q}$, obtain \eqref{eq:exp-A-explicit}.

%To write the resulting expression, define
%\begin{equation}
% \chi_p=1-2p,\qquad
% \dot P_{\nu_j}(\chi_p)=
% \left.\frac{\partial P_\nu(\chi_p)}{\partial\nu}
% \right|_{\nu=\nu_j}.
%\end{equation}

Regarding $B_{jk}$, the two integrals can be obtained by using $R_{jk}$ in \eqref{eq:app-R-solution} and the Laguerre integral identities so that
\begin{equation}\label{eq:app-radial-integrals}
\left\{\begin{aligned}
\mathcal I_{jk}^{R}&=\frac{\Gamma(\nu_j+2)(\nu_j)_k}{k!},\\
\mathcal I_{jk}^{R^2}&=\frac{\Gamma(k+2\nu_j+2)}{k!},
\end{aligned}\right.
\end{equation}
where $(\nu)_0=1$ and $(\nu)_k=\nu(\nu+1)\cdots(\nu+k-1)$ for $k\geq1$. As a result, we can obtain $B_{jk}$ as
\begin{equation}\label{eq:app-radial-weight}
B_{jk}=\frac{\left[\mathcal I_{jk}^{R}\right]^2}{\mathcal I_{jk}^{R^2}}=\frac{\Gamma(\nu_j+2)^2(\nu_j)_k^2}{k!\,\Gamma(k+2\nu_j+2)}.
\end{equation}

For a fixed $j$, we must sum this factor over $k$, including the exponential in \eqref{eq:app-outage-squared-integrals}. To facilitate this calculation, define
\begin{equation}\label{eq:app-radial}
H_\nu(\tau)=\sum_{k=0}^{\infty}\underbrace{\frac{\Gamma(\nu+2)^2(\nu)_k^2}{k!\,\Gamma(k+2\nu+2)}}_{B_{jk}} e^{-2(\nu+k)\tau},
\end{equation}
where the index $j$ is omitted for conciseness. To express \eqref{eq:app-radial} more compactly, use $\Gamma(k+2\nu+2)=\Gamma(2\nu+2)(2\nu+2)_k$ and $e^{-2(\nu+k)\tau}=e^{-2\nu\tau}(e^{-2\tau})^k$. Taking the factors independent of $k$ outside the sum then gives
\begin{equation}
H_\nu(\tau)=\frac{\Gamma(\nu+2)^2}{\Gamma(2\nu+2)} e^{-2\nu\tau}\sum_{k=0}^{\infty}\frac{(\nu)_k^2}{(2\nu+2)_k\,k!}(e^{-2\tau})^k.
\end{equation}
The summation is recognized as the Gauss hypergeometric function with parameters $\nu,\nu,2\nu+2$ and argument $e^{-2\tau}$. As a result, we have obtained \eqref{eq:exp-H} for $H_\nu(\tau)$. Finally, inserting \eqref{eq:app-A-from-integrals} and \eqref{eq:app-radial} into
\eqref{eq:app-outage-squared-integrals} recovers \eqref{eq:exp-spectral}.

\end{document}